\documentclass[]{interact}

\usepackage{epstopdf}%
\usepackage[caption=false]{subfig}%

\usepackage[table,dvipsnames]{xcolor}

\usepackage[hyphens]{url}
\usepackage[colorlinks=true,allcolors=black]{hyperref}
\usepackage[capitalise,noabbrev]{cleveref} %
\usepackage{xspace}
\usepackage[utf8]{inputenc}
\usepackage[T1]{fontenc}
\usepackage{color,soul}

\setul{0.3ex}{0.3ex}
\setulcolor{black}

\usepackage{natbib}%
\bibpunct[, ]{(}{)}{;}{a}{}{,}%
\renewcommand\bibfont{\fontsize{10}{12}\selectfont}%

\theoremstyle{plain}%

\theoremstyle{definition}

\theoremstyle{remark}

\usepackage[acronym]{glossaries}
\usepackage{makecell}
\usepackage{graphicx}

\newacronymstyle{long-short-br}
{%
  \GlsUseAcrEntryDispStyle{long-short}%
}%
{%
  \GlsUseAcrStyleDefs{long-short}%
}
\setacronymstyle{long-short-br}

\newcommand\rev[1]{{#1}} %

\begin{document}

\newcommand{\stargantwo}{StarGAN~v2\xspace}
\newcommand{\stylegantwoada}{StyleGAN2-ADA\xspace}
\newcommand{\stylegantwo}{StyleGAN2\xspace}

\newcommand{\supplementary}{\url{https://github.com/VRI-UFPR/DACov2022}}

\newcommand{\fscore}{F-score\xspace}
\newcommand{\fscores}{F-scores\xspace}

\newacronym{gflops}{GFLOPS}{Giga Floating Point Operations per Second}
\newacronym{mtl}{MTL}{Multi-task Learning}
\newacronym{gan}{GAN}{Generative Adversarial Network}
\newacronym{fcn}{FCN}{Fully Convolutional Network}
\newacronym{gam}{GAM}{Gate Attention Module}
\newacronym{dam}{DAM}{Decoder Attention Module}
\newacronym{rab}{RAB}{Residual Attention Block}
\newacronym{ra}{RA}{Reverse Attention}
\newacronym{idc}{IDC}{Improved Dilation Convolution}
\newacronym{ag}{AG}{Attention Gate}
\newacronym{tasegnet}{TA-SegNet}{Tri-level Attention-based Segmentation Network}
\newacronym{tau}{TAU}{Tri-level Attention Unite}
\newacronym{qapnet}{QAP-Net}{Quadruple Augmented Pyramid Network}
\newacronym{raiunet}{RAIU-Net}{Residual Attention Inception U-Net}
\newacronym{chsnet}{CHS-Net}{Covid-19 Hierarchical Segmentation Network}
\newacronym{csse}{CSSE}{Systems Science and Engineering}
\newacronym{jhu}{JHU}{Johns Hopkins University}
\newacronym{rtpcr}{RT-PCR}{Reverse-Transcription Polymerase Chain Reaction}
\newacronym{ct}{CT}{Computed Tomography}
\newacronym{resnet}{ResNet}{Residual Network}
\newacronym{cnn}{CNN}{Convolutional Neural Network}
\newacronym{relu}{ReLu}{Rectified Linear Units}
\newacronym{flops}{FLOPS}{Floating Point Operations per Second}
\newacronym{senet}{SE-Net}{Squeeze-and-Excitation Networks}
\newacronym{fc}{FC}{Fully Connected}
\newacronym{fpn}{FPN}{Feature Pyramid Network}
\newacronym{psp}{PSPNet}{Pyramid Scene Parsing Network}
\newacronym{ggo}{GGO}{Ground Glass Opacity}
\newacronym{iou}{IoU}{Intersection over Union}
\newacronym{pab}{PAB}{Position-wise Attention Block}
\newacronym{mfab}{MFAB}{Multi-scale Fusion Attention Block}
\newacronym{manet}{MA-Net}{Multi-scale Attention Net}
\newacronym{vgg}{VGG}{Visual Geometry Group Network}
\newacronym{iagan}{IAGAN}{Inception-Augmentation GAN}
\newacronym{randgan}{RANDGAN}{Randomized Generative Adversarial Network}
\newacronym{wss}{WSS}{Weakly-Supervised Segmentation}
\newacronym{stn}{STN}{Spatial Transformer Network}
\newacronym{rnn}{RNN}{Recurrent Neural Network}
\newacronym{gan}{GAN}{Generative Adversarial Network}
\newacronym{adain}{AdaIN}{Adaptive Instance Normalization}
\newacronym{fid}{FID}{Fréchet Inception Distance}
\newacronym{clahe}{CLAHE}{Contrast Limited Adaptive Histogram Equalization}
\newacronym{rbc}{RBC}{Random Brightness Contrast}
\newacronym{da}{DA}{Data Augmentation}
\newacronym{rtpcr}{RT-PCR}{Reverse Transcription–Polymerase Chain Reaction}

\newcommand*{\RL}[2][]{\textcolor{Rhodamine}{[\textbf{\ifthenelse{\equal{#1}{}}{RL}{RL(#1)}}: #2]}}
\newcommand*{\daniel}[2][]{\textcolor{Red}{[\textbf{\ifthenelse{\equal{#1}{}}{Daniel}{Daniel(#1)}}: #2]}}

\title{DACov: A Deeper Analysis of Data Augmentation on the Computed Tomography Segmentation Problem}

\author{\name{Bruno A. Krinski$^*$\thanks{$^*$Corresponding author: Bruno A. Krinski (bakrinski@inf.ufpr.br)\\[1.1ex]This is an author-prepared version. The final published article (the version of record) is available on \textit{Taylor \& Francis Online} (DOI: \href{http://doi.org/10.1080/21681163.2023.2183807}{\textcolor{blue}{10.1080/21681163.2023.2183807}}).}, Daniel V. Ruiz, Rayson Laroca, Eduardo Todt}
\affil{\textsuperscript{Federal University of Paran\'{a}, Curitiba, Brazil} }
}

\maketitle

\begin{abstract}

Due to the COVID-19 global pandemic, computer-assisted diagnoses of medical images have gained much attention, and robust methods of semantic segmentation of \gls*{ct} images have become highly desirable. In this work, we present a deeper analysis of how data augmentation techniques improve segmentation performance on this problem. We evaluate \rev{$20$ traditional augmentation techniques (i.e., not based on neural networks) on five public datasets}. \rev{Six different probabilities of applying each augmentation technique on an image were evaluated}. We also assess a different training methodology where the training subsets are combined into a single larger set. All networks were evaluated through a $5$-fold cross-validation strategy, resulting in over \rev{$4{,}600$} experiments.
We also propose a novel data augmentation technique based on \glspl*{gan} to create new healthy and unhealthy lung \gls*{ct} images, \rev{evaluating four variations of our approach with the same six probabilities of the traditional methods.}
Our findings show that \rev{\gls*{gan}-based techniques and} spatial-level transformations are the most promising for improving the learning of deep models on this problem, \rev{with the \stargantwo + F with a probability of $0.3$ achieving the highest \fscore value on the Ricord1a dataset in the unified training strategy}. Our code is publicly available at \supplementary.

\end{abstract}


\glsresetall
\section{Introduction}
\label{sec:intro}

Since 2019 the world has struggled with the coronavirus pandemic~(COVID-19), with millions of infections and deaths worldwide~\citep{Wang2020}. According to \citet{coviddata} (last updated on 30 Dec. 2022), there are more than $659$M cases and more than $6.6$M deaths globally. Due to the rapid spread of the virus, early diagnosis is highly desirable for faster treatment and screening of infected people~\citep{Chen2020.04.06.20054890,Huang2020}. 

Automatic detection of COVID-19 infections in \gls*{ct} scans has shown to be a great help for early diagnoses~\citep{Shi2021}, with the Semantic Segmentation~\citep{Cao2020} of \gls*{ct} scans with deep learning-based approaches being widely explored since the COVID-19 outbreak~\citep{Shi2021,Narin2021}. Furthermore, performing automatic segmentation of COVID-19 \gls*{ct} images assists doctors in diagnosing and quantifying COVID-19 lesions in a way to avoid human subjectivity~\citep{Zhang2022,Anthimopoulos2016}. Note that the spreading velocity of the virus increases the number of infected and causes a massive shortage of test kits for \gls*{rtpcr}, the primary tool for diagnosing COVID-19. It is also worth mentioning that \gls*{rtpcr} tests have high false negative rates~\citep{Ai2020}. Thus, deep learning methods for COVID-19 \gls*{ct} segmentation have become an essential supplementary tool for \gls*{rtpcr} tests~\citep{Zhang2022}.

However, this process has two main limiting factors. The first one is that labeling images for Semantic Segmentation is a laborious and timing-consuming task, as each image pixel must receive the correct label; to illustrate, for assembling the Cityscapes dataset of urban scenes semantic segmentation~\citep{cordts2016cityscapes}, an average of $90$ minutes was required to label each image. Otherwise, the network may converge to incorrect results~\citep{Shi2021,Cao2020}. The other limiting factor lies in the fact that the annotation of \gls*{ct} segmentation datasets must be made by highly specialized doctors so that the lesion regions of the image are properly labeled~\citep{Shi2021}. 

In this work, we propose an extensive analysis of how data augmentation techniques improve the training of Semantic Segmentation networks on this specific problem. A total of $20$ \rev{ traditional different data augmentation techniques and $4$ that are \gls*{gan} based (variations of our method proposed here) } applied with \rev{six} distinct probabilities were evaluated on five different datasets: MedSeg~\citep{medseg}, Zenodo~\citep{zenodo}, CC-CCII~\citep{Zhang2020}, MosMed~\citep{Morozov2020}, Ricord1a~\citep{ricord1a}, see \cref{sec:dataaugeval}. A unified training strategy, where all training sets were combined, was also evaluated with the $24$ data augmentation techniques (see \cref{sec:trainingunified}). All networks were validated through a $5$-fold cross-validation strategy, thus resulting in over \rev{$4{,}600$} experiments. Additionally, we propose a novel data augmentation technique that exploits a \gls*{gan} to create new healthy and unhealthy lung \gls*{ct} images (see \cref{sec:gandataaug}), \rev{ and we evaluate $4$ variations of our method with the same probabilities of the traditional techniques}. \rev{Our findings show that \glspl*{gan} based techniques and spatial-level transformations are the most promising for improving the learning of neural networks on this~problem, with the \stargantwo + F with a probability $0.3$ achieving the highest \fscore on the Ricord1a dataset in the unified training strategy}. The code for running these same experiments is publicly~available\footnote{Our code is publicly available at~\supplementary}.

A preliminary version of this work was published at the 22th \textit{Simp\'{o}sio Brasileiro de Computa\c{c}\~{a}o Aplicada \`{a} Sa\'{u}de} (Brazilian Symposium on Computing Applied to Health)~\citep{krinski2022}. This work differs from that in several aspects. For this work, we used the original \gls*{ct} image resolution of $512\times512$ instead of $256\times256$ pixels to reduce the information loss generated from the resize operation, which increased the \fscore on four datasets \rev{in comparison with results obtained in previous work~\citep{krinski2022}}; experiments with \rev{six} probabilities \rev{(0.05, 0.1, 0.15, 0.2, 0.25, 0.3)} were conducted and analyzed instead of the previous two (0.1, 0.2); \rev{We chose this range to include the probability $0.15$ previously used on the literature \citep{muller2020automated, Mller2021}.}  We improved the stop criteria for training (i.e., instead of relying on a fixed number of epochs, we adopted the early stopping technique); we present and evaluate a different methodology where the training subsets from different datasets are combined into a single larger set, and we propose a novel technique for synthesizing \gls*{ct} images of lungs with and without lesions. Lastly, here we describe and discuss our experiments in a broader and deeper~manner.
\section{Related Work}

Data augmentation aims to generate synthetic images by applying different operations to preexisting labeled images to aid the learning process of deep learning algorithms in problems that lack available training data~\citep{ruiz2019anda,ruiz2020ida,Zhong2020,chen2020gridmask}, with COVID-19 being one of those problems. However, as COVID-19 \gls*{ct} segmentation is a recent problem, few works proposed data augmentation techniques and, in general, these works are limited to generic data augmentation applications~\citep{Narin2021, Diniz2021, Zhang2022, Salama2022, Yao2022}. The most common operation are variations of an affine transformation, such as random flipping, translation, rotation, and scaling. \citet{muller2020automated, Mller2021} evaluated eight generic data augmentation divided into three categories: spatial, color and noise~transformations. 

Besides generic data augmentation techniques demonstrating promising results, some works prefer to deal with domain adaptation to expand the data available in innovative ways. In \citep{yazdekhasty2021segmentation}, the authors used a conditional \gls*{gan}~\citep{mirza2014conditional} to generate \gls*{ct} images. First, they extracted the lesion regions from the images, generating images with only the lesions. Then, they mirrored these images and fed them into the \gls*{gan} model to generate new synthetic images of lesions. Afterward, the generated lesions replaced the original lesions in the original image. The generated images were used to train a \gls*{fcn} architecture for extracting contextual information from the~images.

\citet{chen2023teacherstudent} also proposed a data augmentation strategy for the COVID-19 \gls*{ct} segmentation problem. They employed the Fourier transformation to convert \gls*{ct} scans from cancer patients to \gls*{ct} scans with COVID-19 lesions. Then, a teacher-student architecture was applied to generate segmentation masks for the new COVID-19 \gls*{ct} scans. The student architecture receives the cancer \gls*{ct} scans converted into COVID-19 \gls*{ct} scans, and a set of unlabeled \gls*{ct} scans with COVID-19 lesions and outputs the segmentation masks. While the student network achieved good results for cancer \gls*{ct} segmentation, good results for COVID-19 were not reached. To overcome this problem, the authors added a teacher architecture trained with unlabeled COVID-19 \gls*{ct} scans to help the student architecture extract robust features of~COVID-19.

In~\citep{Jiang2021}, a data augmentation based on \glspl*{gan} for COVID-19 datasets is proposed. The proposed \gls*{gan} model is designed with the generator and discriminator as dual networks for global and local feature extraction. Also, the generator contains two sub-discriminators to distinguish multi-resolution images. A dynamic element-wise sum process, called DESUM, was proposed to balance the information extracted in the generator step. A dynamic feature matching process, called DFM, was proposed to weight the loss of input with different resolutions~dynamically.

The data augmentation technique proposed by~\citet{mahapatra2021ct} uses geometric information of COVID-19 lesions to generate new samples. In their method, the input image $X$ is fed into a \gls*{wss} module to generate a segmentation mask $S_{X}$. Then, $S_{X}$ inputs a \gls*{stn}~\citep{jaderberg2016spatial} to generate a new segmentation mask changing the shape, location, scale, and orientation of the COVID lesions. The output of the \gls*{stn} model inputs a \gls*{gan}. The generator step of the \gls*{gan} model uses the \gls*{stn} mask to generate a new sample image which inputs the discriminator. The discriminator is composed of two classifier networks to evaluate the accuracy of the class and shape of the new~image.

While there are some explorations in mitigating issues related to the small sizes of COVID-19 datasets (in terms of the number of images)~\citep{muller2020automated, Mller2021, Zhang2022}, an extensive evaluation of the impact of applying various data augmentation techniques on improving the semantic segmentation performance across multiple COVID-19-related datasets is still lacking. We consider that such an assessment can provide a better insight into the generalization and real improvement of deep networks for this task. Therefore, in this work, we present an extensive benchmark of \rev{$20$ traditional data augmentation techniques and $4$ based on \glspl*{gan}, totalling $24$ augmentation techniques. All techniques were applied with six probabilities ($0.05$, $0.1$, $0.15$, $0.2$, $0.25$, and $0.30$)} on five public datasets, using a $5$-fold cross-validation strategy, thus resulting in over \rev{$4{,}600$} experiments, which is the largest benchmark on this field to the best of our~knowledge.
\section{Proposed Work}

In this work, we evaluate $20$ \rev{traditional data augmentation techniques (i.e., not based on neural networks}) on COVID-19~\gls*{ct} scans.
The techniques were applied to an encoder-decoder network composed of RegNetx-002~\citep{xu2021regnet} and U-net++~\citep{Zhou2018}, comparing \rev{six different probabilities of applying the techniques (0.05, 0.1, 0.15, 0.2, 0.25 and 0.30)}.
The input was kept at the original resolution of $512\times512$ pixels --~a higher resolution than previous works such as~\citep{krinski2022}~--; see details in \cref{sec:dataaugeval}.

\begin{figure}[!htb]
\centering
\subfloat{
\resizebox*{14cm}{!}{\includegraphics{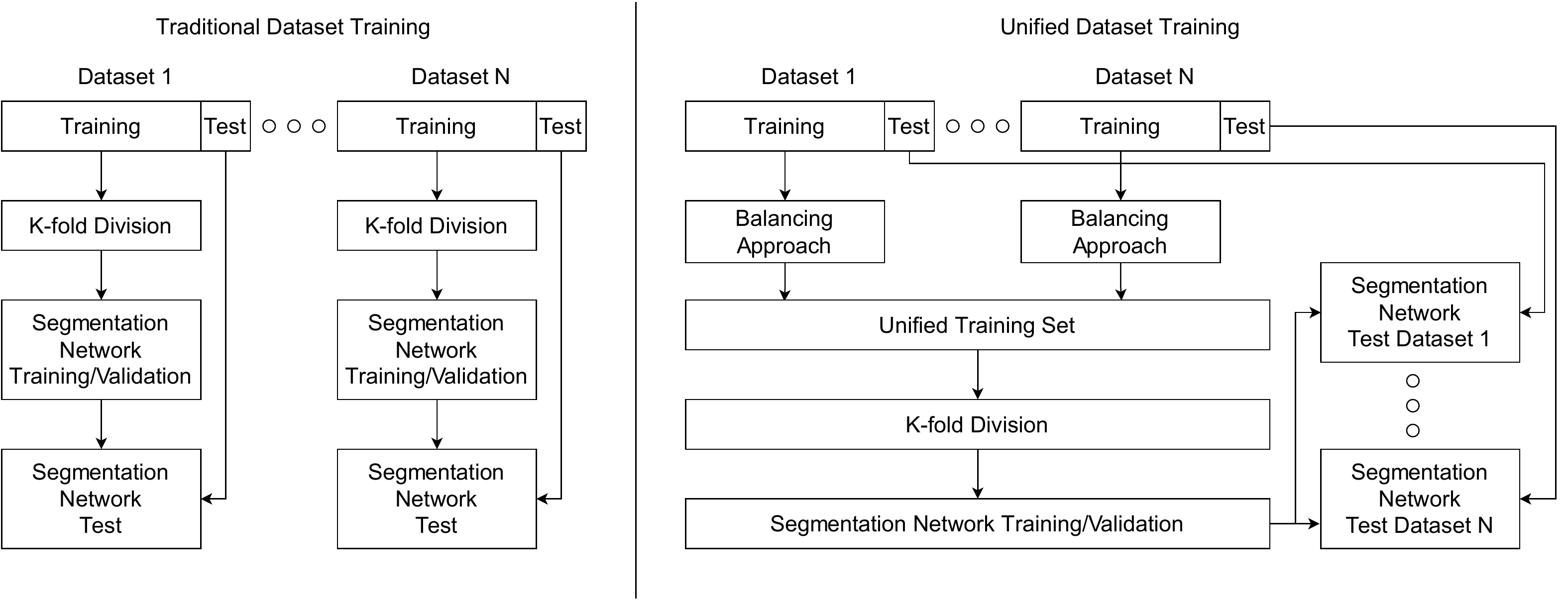}}
}

\vspace{-3mm}

\caption[]{
\rev{Comparison between the traditional training workflow and our unified training methodology. In the traditional approach (left) the datasets are used to train the network in an individual manner, meanwhile, in the unified training strategy (right) the datasets are combined to train the network.}}
\label{fig:unified}
\end{figure}

In addition to the traditional \rev{training} approach, where the training and test sets are disjoint subsets from the same dataset, we propose a different methodology: \rev{as illustrated in \cref{fig:unified} (right), the training subsets are combined into a single larger set while the testing procedure remains essentially the same. %
To ensure a fair comparison, the original classes were adapted to \textbf{background} and \textbf{lesion}. For further details on this methodology, see \cref{sec:trainingunified}.}

\begin{figure}[!htb]
\centering
\subfloat{
\resizebox*{12cm}{!}{\includegraphics{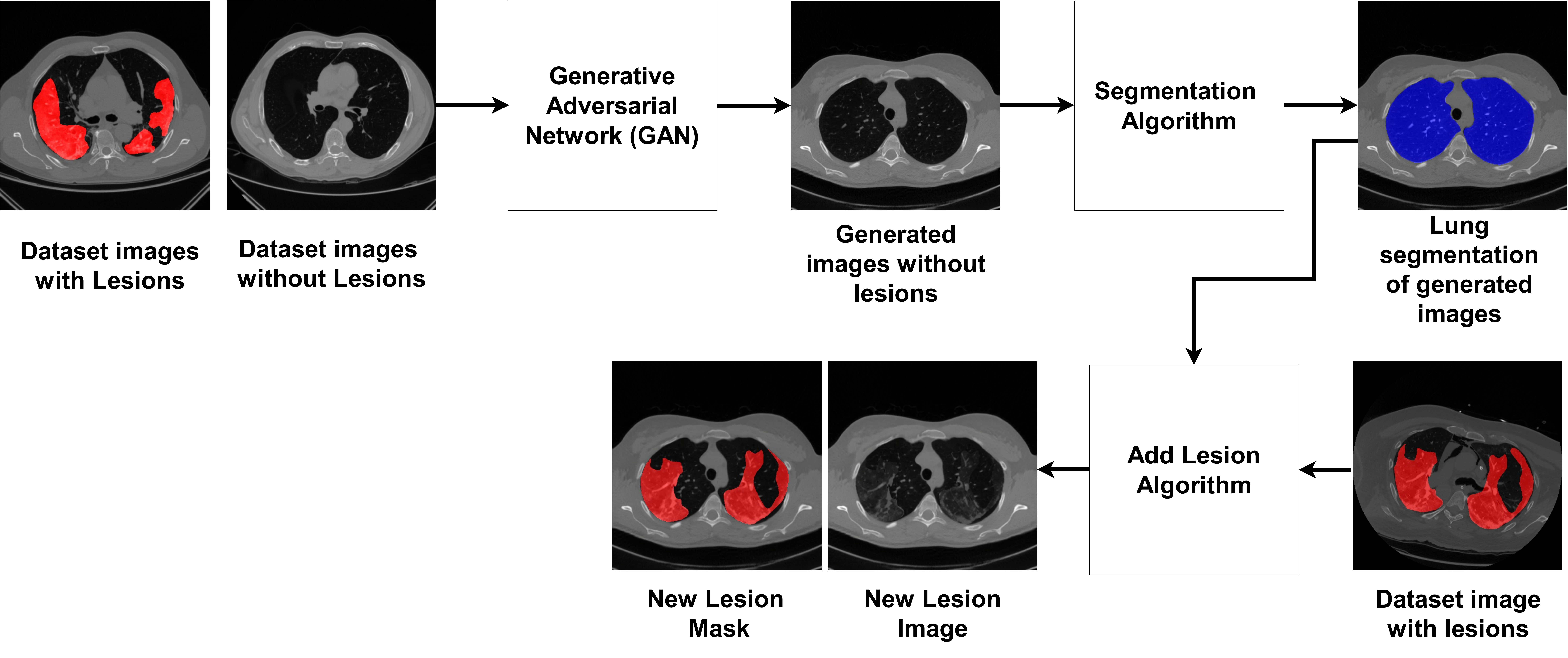}}}

\vspace{-3mm}

\caption[]{\rev{Workflow of our approach, which is divided into three main steps. A \gls*{gan} is trained using two classes, healthy and COVID-19-infected lungs, or only healthy lungs, depending on the network. Thus, the network generates new healthy samples that subsequently have the lung region labeled by a pre-trained segmentation model, ensuring that when the lesions are added later they are cropped by the lung mask. The last step is to add labeled lesions from real \gls*{ct} images into the generated samples. Labeled lesion regions were highlighted in red, labeled lung region was highlighted in blue.}}
\label{fig:pipeline}
\end{figure}

We also propose a novel data augmentation technique that employs a \gls*{gan} model to produce new healthy lungs on \gls*{ct} scans \rev{(see \cref{fig:pipeline})}. These new lungs are combined with preexisting labeled lesions to generate new samples and boost the segmentation training; \rev{ for further details, see \cref{sec:gandataaug}.} \rev{We evaluate four variations of our approach with the same six probabilities of the traditional methods, totaling $24$ different techniques evaluated in this paper.}

\subsection{Datasets and Evaluation Metrics}

The segmentation models were trained and evaluated across five different datasets of \gls*{ct} scans: MedSeg~\citep{medseg}, Zenodo~\citep{zenodo}, CC-CCII~\citep{Zhang2020}, MosMed~\citep{Morozov2020}, and Ricord1a~\citep{ricord1a}. The Ricord1b~\citep{ricord1a} was used to train the \glspl*{gan}. MedSeg has $929$ images and labels for four classes: Background, \gls*{ggo}, Consolidation, and Pleural Effusion. The Zenodo dataset has $3{,}520$ images and labels for four classes: Background, Left Lung, Right Lung, and Infections. The MosMed dataset comprises $2{,}049$ images, with labels for two classes: Background and \gls*{ggo}-Consolidation. The Ricord dataset is divided into 1a, 1b, and 1c. Set 1a is the only one with segmentation masks and has $9{,}166$ images with labels for two classes: Background and Infections. Set 1b has $21{,}220$ \gls*{ct} images with negative diagnostics for COVID-19. We also used a sub-set of CC-CCII with segmentation masks composed of $750$ images containing labels for four classes: Background, Lung Field, \gls*{ggo}, and~Consolidation. \rev{\cref{datasets} summarizes the relevant information about the datasets.}

\begin{table}[!ht]
\centering
\tbl{\rev{The segmentation models were trained and evaluated across five different datasets of \gls*{ct} scans: MedSeg~\citep{medseg}, Zenodo~\citep{zenodo}, CC-CCII~\citep{Zhang2020}, MosMed~\citep{Morozov2020} and Ricord1a~\citep{ricord1a}. The Ricord dataset is divided into 1a, 1b, and 1c. Set 1b has images with negative diagnostics for COVID-19 and was used to train the~\glspl*{gan}.}}
{\resizebox*{\textwidth}{!}{\begin{tabular}{cccccccccccc} \toprule
\multicolumn{1}{c}{\textbf{Dataset}} &
\multicolumn{1}{c}{\textbf{Type}} &
\multicolumn{1}{c}{\textbf{Number of Images}} &
\multicolumn{1}{c}{\textbf{Removed Images}} & 
\multicolumn{1}{c}{\textbf{Labels}} & \\ \midrule
\makecell{CC-CCII} & 
\makecell{Segmentation} & 
\makecell{$750$} & 
\makecell{$201$} & 
\makecell{Background, Lung Field, \\\gls*{ggo}, and~Consolidation} & \\ \midrule
\makecell{MedSeg} & 
\makecell{Segmentation} &
\makecell{$929$} &
\makecell{$457$} &
\makecell{Background, \gls*{ggo}, \\Consolidation, and Pleural Effusion} &\\ \midrule
\makecell{MosMed} &
\makecell{Segmentation} &
\makecell{$2{,}049$} &
\makecell{$1{,}264$} &
\makecell{Background and \\\gls*{ggo}-Consolidation} &\\ \midrule
\makecell{Ricord1a} &  
\makecell{Segmentation} &
\makecell{$9{,}166$} &
\makecell{$0$} &
\makecell{Background and Infections} &\\ \midrule
\makecell{Ricord1b} & 
\makecell{Classification} &
\makecell{$21{,}220$} &
\makecell{$0$} &
\makecell{Positive for COVID-19} &\\ \midrule
\makecell{Zenodo} & 
\makecell{Segmentation} &
\makecell{$3{,}520$} &
\makecell{$1{,}676$} &
\makecell{Background, Left Lung, \\Right Lung, and Infections} \\ \bottomrule
\end{tabular}}}
\label{datasets}
\end{table}

One of the problems pointed out in our previous works~\citep{Krinski2021,krinski2022} was the class imbalance due to several images with just the background class; in fact, recent work has shown that several applications suffer from class imbalance~\citep{johnson2019survey,laroca2021towards, laroca2022cross,sanagavarapu2021covid}. To mitigate this problem,~\citet{krinski2022} removed images with no labeled pixels in the ground-truth mask. However, the CC-CCII and Zenodo datasets remained with masks containing only lung pixels, which still causes an imbalance in the datasets. Therefore, in this work, we removed those images. The total number of removed images in the CC-CCII dataset was $201$, and in the Zenodo dataset was $1{,}676$. Also, in the MedSeg dataset, $457$ images were removed; in the MosMed, $1{,}264$ images were removed; and in the Ricord1a, no image was removed.  \rev{The datasets used in this work do not present predefined subsets for training and testing.
Hence, we divided its data into $80$\% for training and $20$\% for testing following the Pareto principle~\citep{Dunford2014ThePP}. This division is also applied to other segmentation problems~\citep{Urooj2018CVPR, Dmitriev2019, Pandey2020, urcabscale2021, habili2022hyperspectral}, including medical segmentation problems~\citep{dong2018unsupervised, Chen2019, Cao20202, Li2022,Gite2022}}. \rev{In order to perform the training and validation, a standard strategy adopted in the Covid-19 \gls*{ct} segmentation problem is dividing the training set in K-folds, with a $5$-fold cross-validation being the most common division adopted~\citep{Mller2021, Saood2021, yazdekhasty2021segmentation, Fung2021, Sun2022}. Following that, we use a $5$-fold cross-validation strategy avoiding the results being attached to a particular training and validation division}. Following~\citep{Enshaei2022, SAEEDIZADEH2021100007}, the metrics used for evaluation were the \fscore described by~\cref{eq:dice_loss} and \gls*{iou} described by~\cref{eq:jaccard_loss}.

\begin{equation}
    \fscore = \frac{TruePositive}{TruePositive + \frac{FalsePositive + FalseNegative}{2}}
\label{eq:dice_loss}
\end{equation}

\begin{equation}
     IoU = \frac{intersection}{union}
\label{eq:jaccard_loss}
\end{equation}

\rev{The F-score is the harmonic average between Precision and Recall.
Precision evaluates the proportion of pixels classified as positive that are genuinely positive, and Recall evaluates the proportion of positive pixels that were indeed classified as positive~\citep{powers2020evaluation}. True Positive is the number of positive pixels classified as positive (correct classification of positive pixels); False Positive is the number of negative pixels classified as positive (wrong classification of negative pixels); and False Negative is the number of positive pixels classified as negative (wrong classification of positive~pixels).}

\rev{The \gls{iou}~\citep{Jaccard1912} is another metric widely used to evaluate Semantic Segmentation methods~\citep{minaee2022image}. The \gls*{iou} metric takes two areas: the predicted object's area and the target object's area. Then, two values are calculated: these areas' intersection and union.
The intersection is the overlap area of the predicted and target objects, and the union is the sum of both areas. Lastly, the \gls*{iou} is calculated as the intersection value divided by the union value~\citep{minaee2022image}.}

In order to evaluate the \gls*{gan} architectures, the \gls*{fid}~\citep{fid_ref} was used. This metric compared the distribution of  images generated by the \gls*{gan} models with the distribution of real images used to train the discriminator. \gls*{fid} uses one of the deeper layers of the Inception V3 network to compare the mean and standard deviation of the distributions. The lower the \gls*{fid}, the better the \gls*{gan} results. 

The one-sided Wilcoxon signed-rank test was applied to perform a statistical analysis of the data augmentation evaluation. The statistical significance is measured through the P-value, which contains the probability of achieving the measured statistical value when the null hypothesis is true. In order to decide to accept or reject the null hypothesis, a significance level $\alpha = 0.05$ is defined~\citep{statistical_comparisons}. Let $d$ represent the difference between the paired samples: $d = x - y$, where $x$ is the distribution without data augmentation, and $y$ is the distribution with data augmentation. Then, if the p-value is higher than $\alpha$, the null hypothesis is accepted (the underlying distribution $d$ is stochastically higher than a distribution symmetric about zero), or if the P-value is smaller than $\alpha$, the null hypothesis is~rejected.

\rev{The estimated computational cost for the entire encoder-decoder network used here is $22.2$ \gls*{gflops} for an input with a resolution of $512 \times 512$ pixels.
The cost for \stylegantwo according to \citet{liu2021content,xu2022mind} is $45.1$ \gls*{gflops} for $256 \times 256$ with a linear increase for $512 \times 512$ pixels.
The cost for \stargantwo according to \citet{kapoor2021tinystargan} is 120 \gls*{gflops} for $256 \times 256$ also with a linear increase for $512 \times 512$ pixels.
The neural networks used here are implemented using Pytorch~\citep{pytorchNEURIPS2019}.}

\rev{
Note that the hardware used for all of our experiments (detailed in \cref{servers}) is a shared resource among researchers in our department.
}

\begin{table}[!ht]
\centering
\tbl{\rev{Hardware setup used for all of our experiments.}}
{\resizebox*{0.55\textwidth}{!}{\begin{tabular}{cccccccccccc} \toprule
\multicolumn{1}{c}{\textbf{Machine}} &
\multicolumn{1}{c}{\textbf{Memory}} &
\multicolumn{1}{c}{\textbf{GPU}} & \\ \midrule
\makecell{1} & 
\makecell{32GB} & 
\makecell{1 $\times$ NVIDIA TITAN Xp - 12GB} & \\ \midrule
\makecell{2} & 
\makecell{512GB} & 
\makecell{2 $\times$ Tesla P100-SXM2 - 16GB} & \\ \midrule
\makecell{3} & 
\makecell{190GB} & 
\makecell{4 $\times$ Tesla V100-PCIE - 32GB} & \\ \midrule
\makecell{4} & 
\makecell{190GB} &
\makecell{4 $\times$ Tesla V100-PCIE - 32GB} & \\ \bottomrule
\end{tabular}}}
\label{servers}
\end{table}

\subsection{Experiment I: Traditional Data Augmentation Techniques}
\label{sec:dataaugeval}

\begin{figure}[!htb]
\centering
\subfloat[][]{
	    \includegraphics[width=0.15\textwidth]{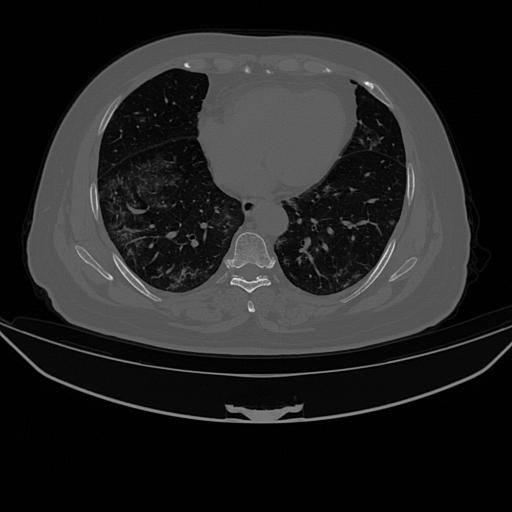}
	    \label{fig:original_image_das}
	}
	
\subfloat[][]{
	    \includegraphics[width=0.15\textwidth]{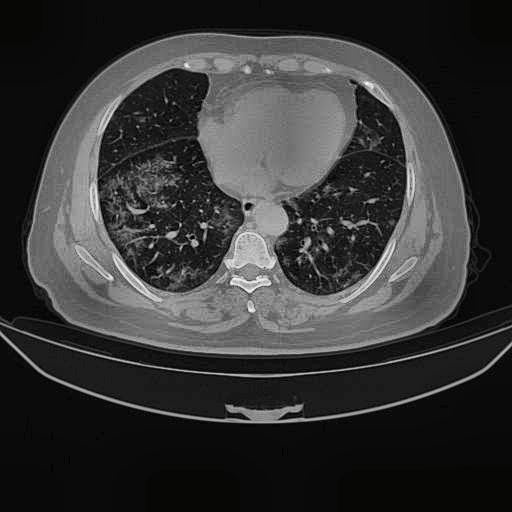}
	    \label{fig:clahe_das}
	}
\subfloat[][]{
	    \includegraphics[width=0.15\textwidth]{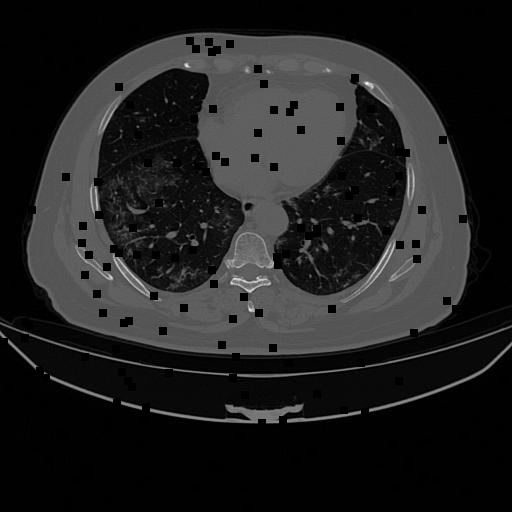}
	    \label{fig:coarse_dropout_das}
	}
\subfloat[][]{
	    \includegraphics[width=0.15\textwidth]{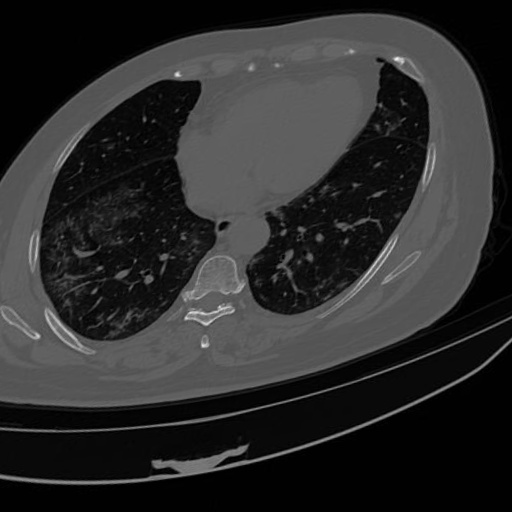}
	    \label{fig:elastic_transform_das}
	}
\subfloat[][]{
	    \includegraphics[width=0.15\textwidth]{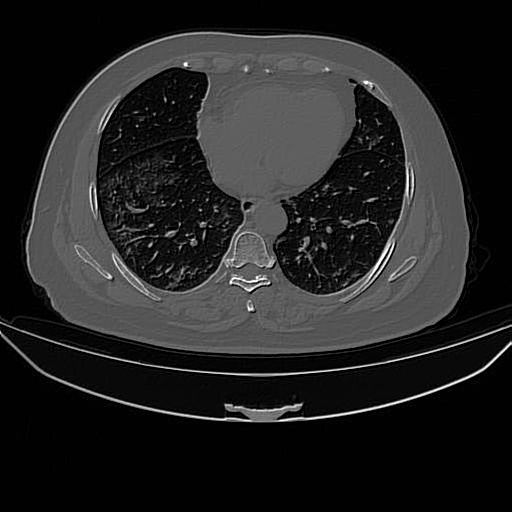}
	    \label{fig:emboss_das}
	}
\subfloat[][]{
	    \includegraphics[width=0.15\textwidth]{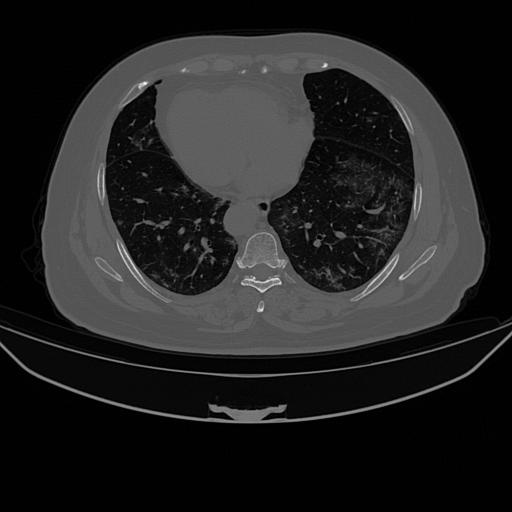}
	    \label{fig:flip_das}
	}
	
\subfloat[][]{
	    \includegraphics[width=0.15\textwidth]{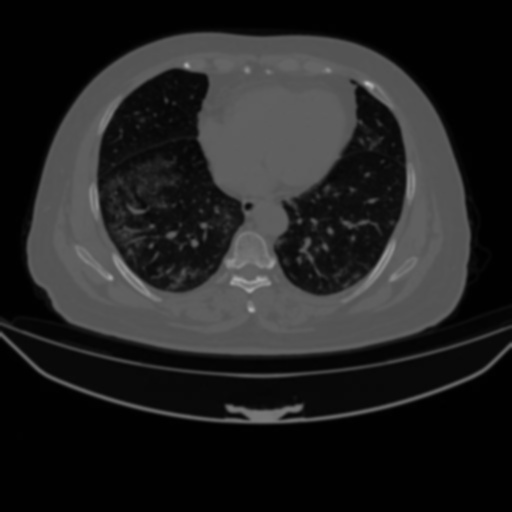}
	    \label{fig:gaussian_blur_das}
	}
\subfloat[][]{
	    \includegraphics[width=0.15\textwidth]{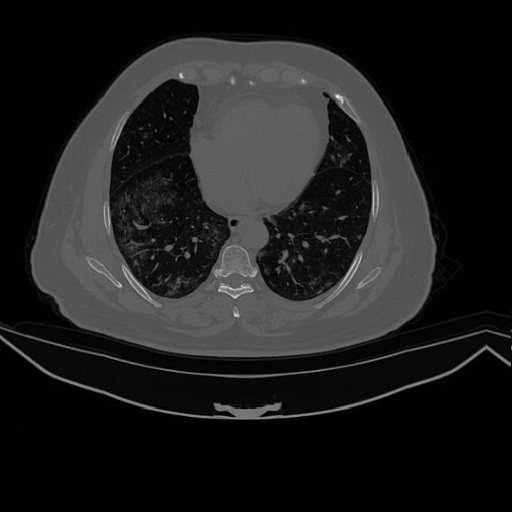}
	    \label{fig:grid_distortion_das}
	}
\subfloat[][]{
	    \includegraphics[width=0.15\textwidth]{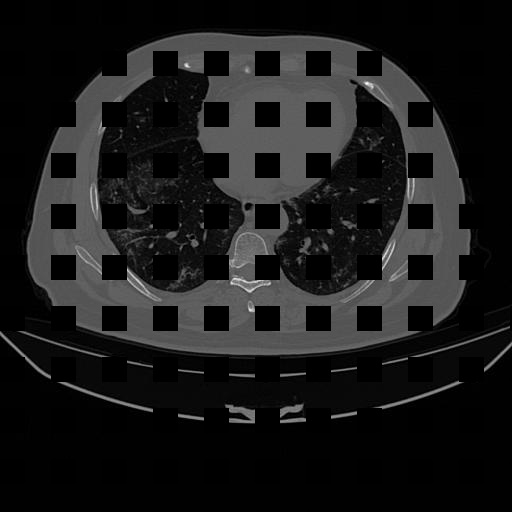}
	    \label{fig:grid_dropout_das}
	}
\subfloat[][]{
	    \includegraphics[width=0.15\textwidth]{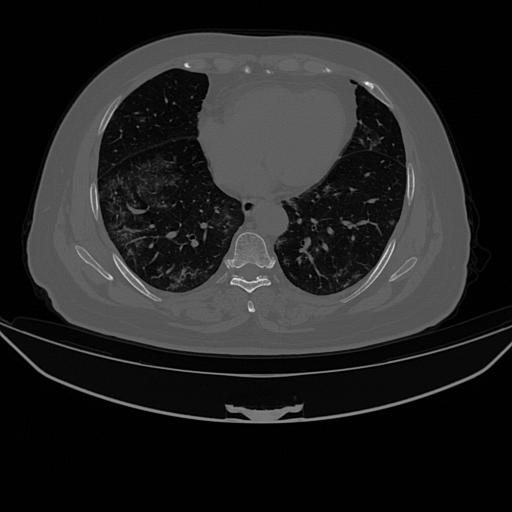}
	    \label{fig:image_Compression_das}
	}
\subfloat[][]{
	    \includegraphics[width=0.15\textwidth]{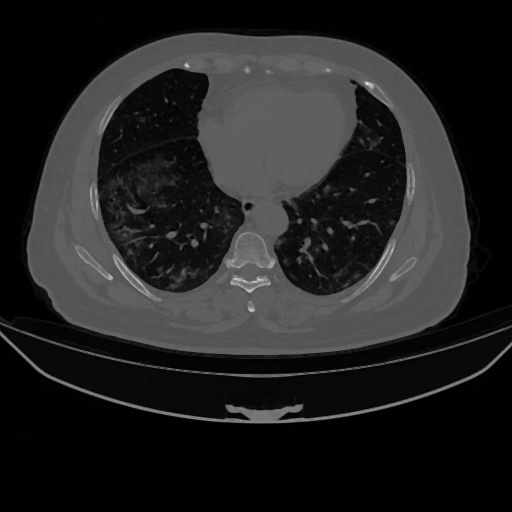}
	    \label{fig:median_blur_das}
	}
	
\subfloat[][]{
	    \includegraphics[width=0.15\textwidth]{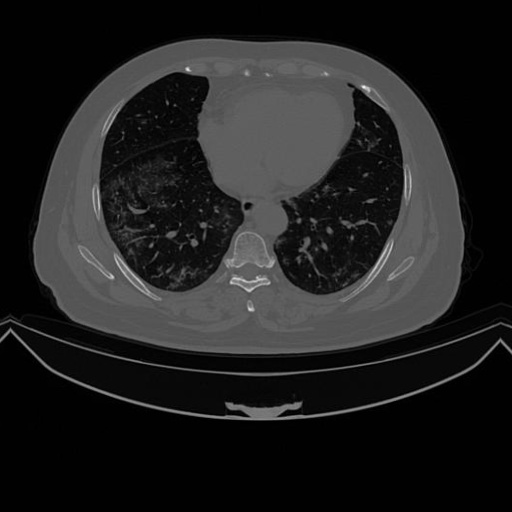}
	    \label{fig:optical_distortion_das}
	}
\subfloat[][]{
	    \includegraphics[width=0.15\textwidth]{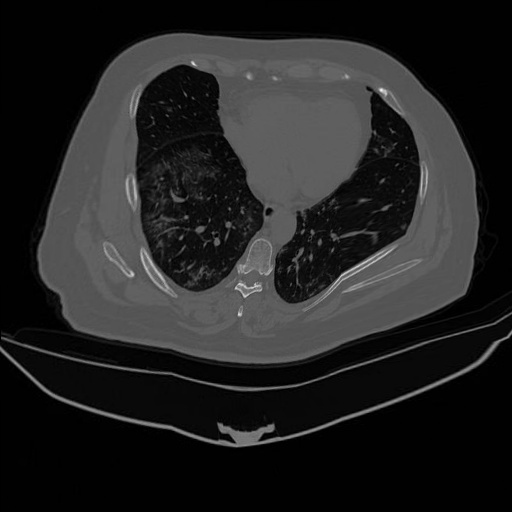}
	    \label{fig:piecewise_affine_das}
	}
\subfloat[][]{
	    \includegraphics[width=0.15\textwidth]{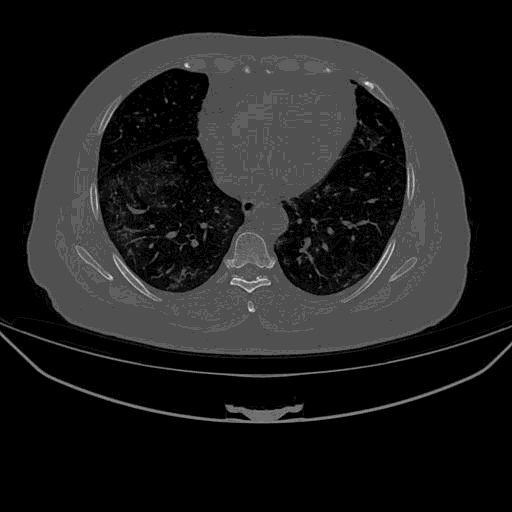}
	    \label{fig:posterize_das}
	}
\subfloat[][]{
	    \includegraphics[width=0.15\textwidth]{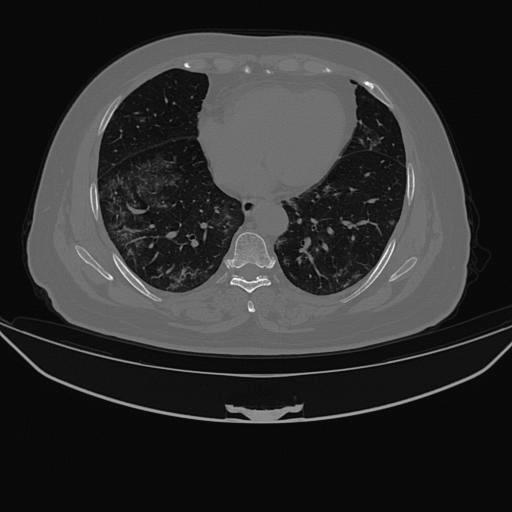}
	    \label{fig:random_brightness_contrast_das}
	}
\subfloat[][]{
	    \includegraphics[width=0.15\textwidth]{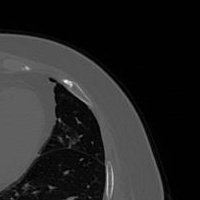}
	    \label{fig:random_crop_das}
	}
	
\subfloat[][]{
	    \includegraphics[width=0.15\textwidth]{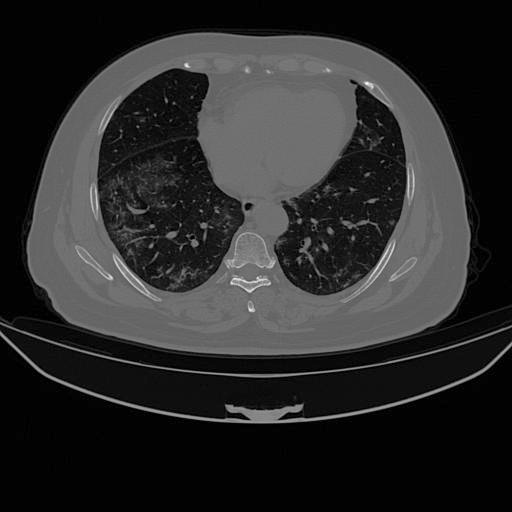}
	    \label{fig:random_gamma_das}
	}
\subfloat[][]{
	    \includegraphics[width=0.15\textwidth]{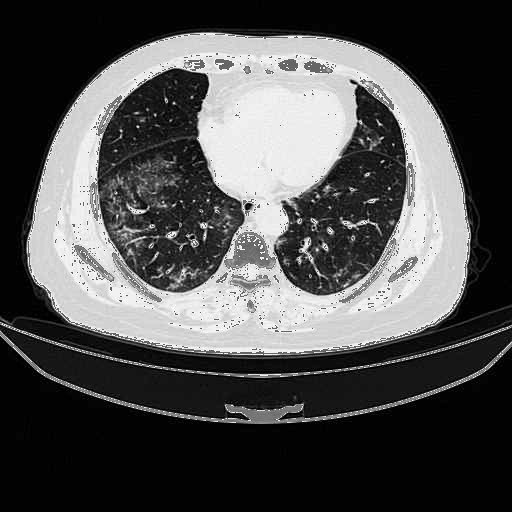}
	    \label{fig:random_snow_das}
	}
\subfloat[][]{
	    \includegraphics[width=0.15\textwidth]{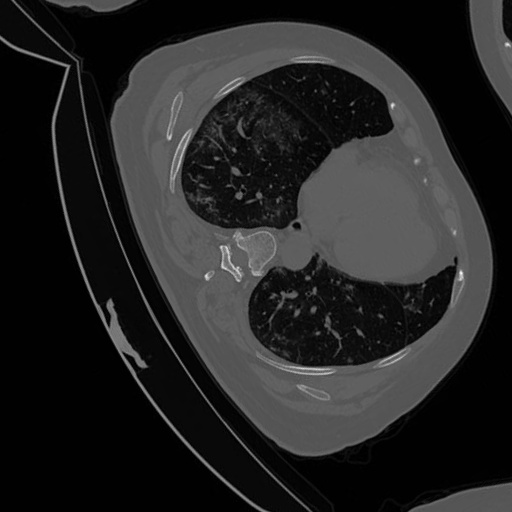}
	    \label{fig:rotate_das}
	}
\subfloat[][]{
	    \includegraphics[width=0.15\textwidth]{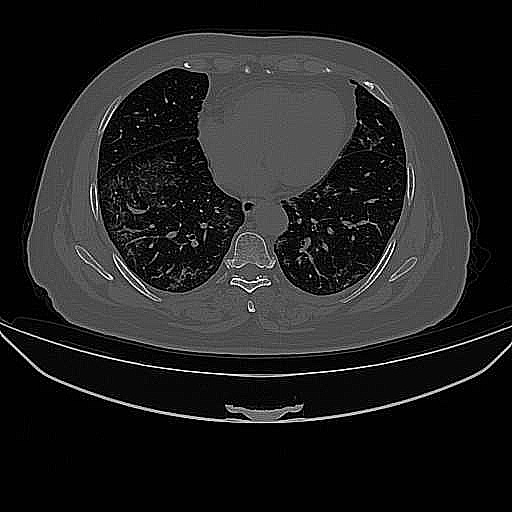}
	    \label{fig:sharpen_das}
	}
\subfloat[][]{
	    \includegraphics[width=0.15\textwidth]{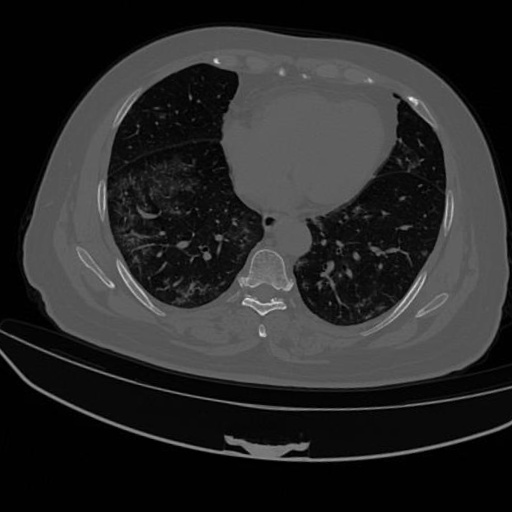}
	    \label{fig:shift_scale_rotate_das}
	}

\caption{\rev{Illustration of the $20$ traditional data augmentation techniques (i.e., not based on neural networks) applied to a \gls*{ct} image. In~\ref{fig:original_image_das}, the original image is presented. %
The data augmentation techniques are illustrated in the following sequence: \gls{clahe}~\ref{fig:clahe_das}, Coarse Dropout~\ref{fig:coarse_dropout_das}, Elastic Transform~\ref{fig:elastic_transform_das}, Emboss~\ref{fig:emboss_das}, Flip~\ref{fig:flip_das}, Gaussian Blur~\ref{fig:gaussian_blur_das}, Grid Distortion~\ref{fig:grid_distortion_das}, Grid Dropout~\ref{fig:grid_dropout_das}, Image Compression~\ref{fig:image_Compression_das}, Median Blur~\ref{fig:median_blur_das}, Optical Distortion~\ref{fig:optical_distortion_das}, Piecewise Affine~\ref{fig:piecewise_affine_das}, Posterize~\ref{fig:posterize_das}, \gls{rbc}~\ref{fig:random_brightness_contrast_das}, Random Crop~\ref{fig:random_crop_das}, Random Gamma~\ref{fig:random_gamma_das}, Random Snow~\ref{fig:random_snow_das}, Rotate~\ref{fig:rotate_das}, Sharpen~\ref{fig:sharpen_das}, and Shift Scale Rotate~\ref{fig:shift_scale_rotate_das}.}
}
\label{fig:das}
\end{figure}

\begin{table}[!ht]
\centering
\tbl{Results of the data augmentation evaluation. $p$ stands for probability, the blue-colored values indicate the best \fscore values, while the red-colored values indicate the best \gls*{iou} values. The  values highlighted in green show the data augmentation techniques in which the P-value achieved values lower than $0.05$, and thus the null hypothesis was rejected (i.e., there is a statistical difference and the results achieved are better than without data augmentation).}
{\resizebox*{\textwidth}{!}{
\begin{tabular}{cccccccccccc} \toprule
\multicolumn{1}{c}{\textbf{$p$}} &
\multicolumn{1}{c}{\textbf{Augmentation}} &
\multicolumn{2}{c}{\textbf{CC-CCII}} &
\multicolumn{2}{c}{\textbf{MedSeg}} &
\multicolumn{2}{c}{\textbf{MosMed}} &
\multicolumn{2}{c}{\textbf{Ricord1a}} &
\multicolumn{2}{c}{\textbf{Zenodo}} \\ \midrule
\makecell{} &
\makecell{} &
\makecell{\fscore} &
\makecell{\gls*{iou}} & 
\makecell{\fscore} &
\makecell{\gls*{iou}} & 
\makecell{\fscore} &
\makecell{\gls*{iou}} & 
\makecell{\fscore} &
\makecell{\gls*{iou}} & 
\makecell{\fscore} &
\makecell{\gls*{iou}} \\ \midrule
\makecell{} &
\makecell{No Augmentation}&
\makecell{0.8450}&
\makecell{0.7896}&
\makecell{0.8885}&
\makecell{0.8261}&
\makecell{0.8180}&
\makecell{0.7528}&
\makecell{0.8875}&
\makecell{0.8275}&
\makecell{0.9072}&
\makecell{0.8490} \\
\bottomrule
\end{tabular}}}
\label{results1}
\end{table}

In general, the impacts of \gls*{da} techniques are not approached in studies proposed for the COVID-19 \gls*{ct} segmentation problem~\citep{Saood2021, Mahmud2021}. While some studies apply some data augmentation techniques~\citep{zhao2021d2a,qiblawey2021detection,JosephRaj2021,muller2020automated,Mller2021,chen2020residual,xu2020gasnet}, most of them are limited to flip and rotation transformations. To correctly measure the impact of data augmentation on the COVID-19 \gls*{ct} segmentation problem, we evaluate $20$ different data augmentation techniques \rev{not based on neural networks}: \gls*{clahe}, Coarse Dropout, Elastic Transform, Emboss, Flip, Gaussian Blur, Grid Distortion, Grid Dropout, Image Compression, Median Blur, Optical Distortion, Piecewise Affine, Posterize, \gls*{rbc}, Random Crop, Random Gamma, Random Snow, Rotate, Sharpen, Shift Scale Rotate. \rev{\cref{fig:das} illustrates the 20 data augmentation techniques applied to a \gls*{ct} image.} \rev{These data augmentation techniques were applied --~with the default parameters~-- using the Albumentations library~\citep{info11020125}, which has been successfully explored in various areas of computer vision~\citep{kupyn2019deblurganv2,kaissis2021endtoend,laroca2022first}. Parameter optimization of the best techniques was left for future~work.}

\begin{table}[!ht]
\centering
\tbl{Results of the data augmentation evaluation (Continuation of \cref{results1}). $p$ stands for probability, the blue-colored values indicate the best \fscore values, while the red-colored values indicate the best \gls*{iou} values. The  values highlighted in green show the data augmentation techniques where the P-value achieved values lower than $0.05$, and thus the null hypothesis was rejected (i.e., there is a statistical difference and the results achieved are better than without data augmentation).}
{\resizebox*{\textwidth}{!}{
\begin{tabular}{cccccccccccc} \toprule
\multicolumn{1}{c}{\textbf{$p$}} &
\multicolumn{1}{c}{\textbf{Augmentation}} &
\multicolumn{2}{c}{\textbf{CC-CCII}} &
\multicolumn{2}{c}{\textbf{MedSeg}} &
\multicolumn{2}{c}{\textbf{MosMed}} &
\multicolumn{2}{c}{\textbf{Ricord1a}} &
\multicolumn{2}{c}{\textbf{Zenodo}} \\ \midrule
\makecell{} &
\makecell{} &
\makecell{\fscore} &
\makecell{\gls*{iou}} & 
\makecell{\fscore} &
\makecell{\gls*{iou}} & 
\makecell{\fscore} &
\makecell{\gls*{iou}} & 
\makecell{\fscore} &
\makecell{\gls*{iou}} & 
\makecell{\fscore} &
\makecell{\gls*{iou}} \\ \midrule
\makecell{} &
\makecell{No Augmentation}&
\makecell{0.8450}&
\makecell{0.7896}&
\makecell{0.8885}&
\makecell{0.8261}&
\makecell{0.8180}&
\makecell{0.7528}&
\makecell{0.8875}&
\makecell{0.8275}&
\makecell{0.9072}&
\makecell{0.8490} \\
\bottomrule
\end{tabular}}}
\label{results2}
\end{table}

The encoder-decoder network chosen to evaluate the augmentation techniques was the RegNetx-002~\citep{xu2021regnet} encoder and U-net++~\citep{Zhou2018} decoder. All experiments were evaluated through a $5$-fold cross-validation strategy. We trained the chosen architecture for a maximum of $100$ epochs, with patience~=~$10$ (the number of epochs with no improvement after which training is stopped). The initial learning rate was $0.001$ and was divided by $10$ every $10$ epochs. The augmentation algorithms were applied through a technique called online augmentation. In this technique, the data augmentation is applied during the network training, and each batch of images has a probability of suffering an augmentation operation before inputting the network. Differently from our previous work~\citep{krinski2022}, here we also analyze the effects of increasing the number of images augmented, with \rev{six} probabilities of applying the data augmentation being evaluated: \rev{$0.05$, $0.1$, $0.15$, $0.2$, $0.25$, $0.30$}. 

\rev{In order to avoid information loss when the image is downscaled, the segmentation network was trained with the original resolution of $512\times512$ instead of the downscaled $256\times256$ resolution used in our previous work~\citep{krinski2022}. Downscaling the image generates a loss of information in the images, which negatively affects network learning. Compared with~\citep{krinski2022}, training without downscaling generated impressive gains in \fscore in four of the five evaluated datasets. The most significant difference was in the MedSeg dataset with the Shift Scale Rotate augmentation (the best results achieved), where the results increased from $0.4806$ with an image size of $256\times256$ to $0.8890$ with an image size of $512\times512$, a gain of $0.4084$ in the \fscore~value.} %

\cref{results1} presents the evaluation results for probabilities $0.05$ and $0.1$, \cref{results2} presents the results for probabilities $0.15$ and $0.2$, \rev{and \cref{results5} presents the results for probabilities $0.25$ and $0.3$}. The values highlighted in green show the data augmentation techniques where the P-value achieved values lower than $0.05$, and the null hypothesis was rejected (i.e., there is a statistical difference and the results achieved are better than without data augmentation).

\begin{table}[!ht]
\centering
\tbl{\rev{Results of the data augmentation evaluation (Continuation of \cref{results1} and \cref{results2}). $p$ stands for probability, the blue-colored values indicate the best \fscore values, while the red-colored values indicate the best \gls*{iou} values. The  values highlighted in green show the data augmentation techniques in which the P-value achieved values lower than $0.05$, and thus the null hypothesis was rejected (i.e., there is a statistical difference and the results achieved are better than without data augmentation).}}
{\resizebox*{\textwidth}{!}{
\begin{tabular}{cccccccccccc} \toprule
\multicolumn{1}{c}{\textbf{$p$}} &
\multicolumn{1}{c}{\textbf{Augmentation}} &
\multicolumn{2}{c}{\textbf{CC-CCII}} &
\multicolumn{2}{c}{\textbf{MedSeg}} &
\multicolumn{2}{c}{\textbf{MosMed}} &
\multicolumn{2}{c}{\textbf{Ricord1a}} &
\multicolumn{2}{c}{\textbf{Zenodo}} \\ \midrule
\makecell{} &
\makecell{} &
\makecell{\fscore} &
\makecell{\gls*{iou}} & 
\makecell{\fscore} &
\makecell{\gls*{iou}} & 
\makecell{\fscore} &
\makecell{\gls*{iou}} & 
\makecell{\fscore} &
\makecell{\gls*{iou}} & 
\makecell{\fscore} &
\makecell{\gls*{iou}} \\ \midrule
\makecell{} &
\makecell{No Augmentation}&
\makecell{0.8450}&
\makecell{0.7896}&
\makecell{0.8885}&
\makecell{0.8261}&
\makecell{0.8180}&
\makecell{0.7528}&
\makecell{0.8875}&
\makecell{0.8275}&
\makecell{0.9072}&
\makecell{0.8490} \\
\bottomrule
\end{tabular}}}
\label{results5}
\end{table}

Most data augmentation techniques did not improve the \fscore and the \gls*{iou} values. For the probability of $0.05$, no method achieved an \fscore improvement higher than $1$\% compared to the baseline, where no data augmentation technique was applied. However, seven techniques achieved statistical difference in the Ricord1a, and one technique achieved statistical differences in CC-CCII and Zenodo. For probability $0.1$, Piecewise Affine and Rotate improved the \fscore in the CC-CCII; Shift Scale Rotate improved the \fscore on the MedSeg, and Grid Distortion improved the \fscore on the MosMed. With a probability of $0.15$, Optical Distortion improved the \fscore on the CC-CCII. With a probability of $0.2$, $6$ data augmentation techniques improved the \fscore on the CC-CCII, and $5$ techniques improved the \fscore on the MosMed. Elastic Transform improved the \fscore on the Zenodo dataset. \rev{With a probability of $0.25$, $3$ data augmentation techniques reached better \fscore values on CC-CCII, and $3$ techniques achieved better \fscore values on MosMed.
With a probability of $0.3$, $7$ data augmentation techniques improved the \fscore value obtained on CC-CCII, $5$ techniques improved the \fscore achieved on MosMed, and $1$ technique (Gaussian Blur) improved the \fscore reached on Ricord1a. In general, the probability of $0.3$ produced better results considering the number of techniques that improved the \fscore values.}

\rev{In summary, the best result per dataset was: CC-CCII with Shift Scale Rotate applied with a probability of $0.3$, achieving an \fscore of $0.8567$; MedSeg with Shift Scale Rotate applied with a probability of $0.1$, reaching an \fscore of $0.8905$; MosMed with Flip applied with a probability of $0.25$, obtaining an \fscore of $0.8284$; Ricord1a with Gaussian Blur applied with a probability of $0.3$, achieving an \fscore of $0.8903$; and Zenodo with Elastic Transform applied with a probability of $0.2$, attaining an \fscore of $0.91$.
Thus, in the CC-CCII, MedSeg, MosMed and Zenodo datasets the highest \fscore was achieved through a spatial transformation, whereas in Ricord1a it was reached with a color~operation.}

\rev{In four datasets, the best augmentation techniques are spatial-based operations.
Such techniques provided better results in those datasets because they encourage shape variation of the lesion regions. Unlike other datasets, Ricord1a has very similar images, with minimal changes in lung position and shape. Therefore, applying a spatial operation in this dataset is counterproductive. As a result, a color operation (Gaussian Blur) achieved the highest \fscore, indicating that this dataset is more sensitive to color~operations.}

\subsection{Proposed Methodology: Training Sets Unified}
\label{sec:trainingunified}

In addition to the traditional approach, where the training and test sets are disjoint subsets from the same dataset, we employ a different methodology that combines the training subset from each dataset into a single larger set \rev{(See \cref{fig:unified})}. The original classes of the five datasets were rearranged into two classes: \textbf{background} and \textbf{lesion}, where everything that was not a type of lesion was converted to the background, and all lesions’ sub-types were merged into a single class. The reasoning is to encourage generalization since the trained model is expected to perform reasonably well with any \gls*{ct} image of the lungs.

Besides the rearrangement of the classes, an additional balancing procedure was employed to ensure a fair representation of each original sample on the new combined training set. In summary, smaller datasets (in terms of the number of images), i.e., CC-CCII, MedSeg, MosMed, and Zenodo, had their samples repeated $n$ times to match the number of samples of the largest dataset Ricord1a. For each dataset $i$, $n_i$ was calculated as the ceil of the number of the images in the Ricord1a dataset divided by the number of images in the corresponding dataset $x_i$, as described by \cref{eq:sizeset}.

\begin{equation}
\label{eq:sizeset}
    n_i = \left \lceil{\frac{|Ricord1a|}{|x_i|}} \right \rceil
\end{equation}

\subsubsection{Unified Set Evaluation}

\begin{table}[!ht]
\centering
\tbl{Results of the data augmentation evaluation when unifying the training sets. $p$ stands for probability, the blue-colored values indicate the best \fscore value, and the red-colored values indicate the best \gls*{iou} values. The values highlighted in green show the data augmentation techniques in which the P-value achieved values lower than $0.05$, and thus the null hypothesis was rejected (i.e., there is a statistical difference and the results achieved are better than without data augmentation). The underscored values show the techniques where training with the unified set achieved a P-value lower than $0.05$ when compared with training with a single training set, and the null hypothesis was~rejected.}
{\resizebox*{\textwidth}{!}{
\begin{tabular}{cccccccccccc} \toprule
\multicolumn{1}{c}{\textbf{$p$}} &
\multicolumn{1}{c}{\textbf{Augmentation}} &
\multicolumn{2}{c}{\textbf{CC-CCII}} &
\multicolumn{2}{c}{\textbf{MedSeg}} &
\multicolumn{2}{c}{\textbf{MosMed}} &
\multicolumn{2}{c}{\textbf{Ricord1a}} &
\multicolumn{2}{c}{\textbf{Zenodo}} \\ \midrule
\makecell{} &
\makecell{} &
\makecell{\fscore} &
\makecell{\gls*{iou}} & 
\makecell{\fscore} &
\makecell{\gls*{iou}} & 
\makecell{\fscore} &
\makecell{\gls*{iou}} & 
\makecell{\fscore} &
\makecell{\gls*{iou}} & 
\makecell{\fscore} &
\makecell{\gls*{iou}} \\ \midrule
\makecell{} &
\makecell{No Augmentation}&
\makecell{\ul{0.8636}}&
\makecell{0.8087}&
\makecell{0.8881}&
\makecell{0.8253}&
\makecell{0.8185}&
\makecell{0.7547}&
\makecell{0.8599}&
\makecell{0.7947}&
\makecell{0.9096}&
\makecell{0.8514} \\
\bottomrule
\end{tabular}}}
\label{results3}
\end{table}

\cref{results3} presents the evaluation results for probabilities $0.05$ and $0.1$, \cref{results4} presents the results for probabilities $0.15$ and $0.2$, \rev{and \cref{results6} presents the results for probabilities $0.25$ and $0.3$}. The values highlighted in green show the data augmentations where the P-value achieved values lower than $0.05$, and the null hypothesis was rejected (i.e., there is a statistical difference and the results achieved are better than without data augmentation). Training with a unified training set was also compared with training with individual training sets. The underscored values presented in \cref{results3}, \cref{results4} \rev{and \cref{results6}} show the techniques in which training with the unified set achieved a P-value lower than $0.05$, and the null hypothesis was rejected (i.e., there is a statistical difference and the results achieved in the unified training set are better than the individual training sets).

As presented in \cref{results3}, \cref{results4} \rev{and \cref{results6}}, using the unified training set achieved promising results compared to using the individual training sets. For a probability of $0.05$, training with the unified training set achieved a higher \fscore when compared with the baseline in all data augmentation techniques applied in the CC-CCII dataset. The same occurred in $11$ data augmentation techniques in MosMed and $10$ data in~Zenodo.

\begin{table}[!ht]
\centering
\tbl{Results of the data augmentation evaluation when unifying the training sets (Continuation of \cref{results3}). $p$ stands for probability, the blue-colored values indicate the best \fscore values, and the red-colored values indicate the best \gls*{iou} values. The values highlighted in green show the data augmentation techniques in which the P-value achieved values lower than $0.05$, and thus the null hypothesis was rejected (i.e., there is a statistical difference and the results achieved are better than without data augmentation). The underscored values show the techniques where training with the combined training sets achieved a P-value lower than $0.05$ when compared with training with a single training set, and the null hypothesis was~rejected.}
{\resizebox*{\textwidth}{!}{
\begin{tabular}{cccccccccccc} \toprule
\multicolumn{1}{c}{\textbf{$p$}} &
\multicolumn{1}{c}{\textbf{Augmentation}} &
\multicolumn{2}{c}{\textbf{CC-CCII}} &
\multicolumn{2}{c}{\textbf{MedSeg}} &
\multicolumn{2}{c}{\textbf{MosMed}} &
\multicolumn{2}{c}{\textbf{Ricord1a}} &
\multicolumn{2}{c}{\textbf{Zenodo}} \\ \midrule
\makecell{} &
\makecell{} &
\makecell{\fscore} &
\makecell{\gls*{iou}} & 
\makecell{\fscore} &
\makecell{\gls*{iou}} & 
\makecell{\fscore} &
\makecell{\gls*{iou}} & 
\makecell{\fscore} &
\makecell{\gls*{iou}} & 
\makecell{\fscore} &
\makecell{\gls*{iou}} \\ \midrule
\makecell{} &
\makecell{No Augmentation}&
\makecell{\ul{0.8636}}&
\makecell{0.8087}&
\makecell{0.8881}&
\makecell{0.8253}&
\makecell{0.8185}&
\makecell{0.7547}&
\makecell{0.8599}&
\makecell{0.7947}&
\makecell{0.9096}&
\makecell{0.8514} \\
\bottomrule
\end{tabular}}}
\label{results4}
\end{table}

\begin{table}[!ht]
\centering
\tbl{\rev{Results of the data augmentation evaluation when unifying the training sets (Continuation of \cref{results3} and \cref{results4}). $p$ stands for probability, the blue-colored values indicate the best \fscore values, and the red-colored values indicate the best \gls*{iou} values. The values highlighted in green show the data augmentation techniques in which the P-value achieved values lower than $0.05$, and thus the null hypothesis was rejected (i.e., there is a statistical difference and the results achieved are better than without data augmentation). The underscored values show the techniques in which training with the combined training sets achieved a P-value lower than $0.05$ when compared with training with a single training set, and the null hypothesis was rejected.}}
{\resizebox*{\textwidth}{!}{
\begin{tabular}{cccccccccccc} \toprule
\multicolumn{1}{c}{\textbf{$p$}} &
\multicolumn{1}{c}{\textbf{Augmentation}} &
\multicolumn{2}{c}{\textbf{CC-CCII}} &
\multicolumn{2}{c}{\textbf{MedSeg}} &
\multicolumn{2}{c}{\textbf{MosMed}} &
\multicolumn{2}{c}{\textbf{Ricord1a}} &
\multicolumn{2}{c}{\textbf{Zenodo}} \\ \midrule
\makecell{} &
\makecell{} &
\makecell{\fscore} &
\makecell{\gls*{iou}} & 
\makecell{\fscore} &
\makecell{\gls*{iou}} & 
\makecell{\fscore} &
\makecell{\gls*{iou}} & 
\makecell{\fscore} &
\makecell{\gls*{iou}} & 
\makecell{\fscore} &
\makecell{\gls*{iou}} \\ \midrule
\makecell{} &
\makecell{No Augmentation}&
\makecell{\ul{0.8636}}&
\makecell{0.8087}&
\makecell{0.8881}&
\makecell{0.8253}&
\makecell{0.8185}&
\makecell{0.7547}&
\makecell{0.8599}&
\makecell{0.7947}&
\makecell{0.9096}&
\makecell{0.8514} \\
\bottomrule
\end{tabular}}}
\label{results6}
\end{table}

For a probability of $0.1$, the unified training set achieved a better \fscore value in $2$ data augmentation techniques in the MedSeg dataset, $13$ in MosMed, $11$ in Zenodo, and all data augmentation techniques in the CC-CCII dataset. For a probability of $0.15$, the unified training set achieved a better \fscore value in $6$ data augmentation techniques in the MedSeg dataset, $17$ in MosMed, $13$ in Zenodo, and $19$ in CC-CCII. For a probability of $0.2$, the unified training set achieved better \fscore values in $8$ data augmentation techniques in MedSeg, $12$ in MosMed, $11$ in Zenodo, and all data augmentation techniques in the CC-CCII~dataset.

\rev{For a probability of $0.25$, the unified training set achieved a better \fscore value in $9$ data augmentation techniques in the MedSeg dataset, $17$ in MosMed, $15$ in Zenodo. Finally, for a probability of $0.3$, the unified training set achieved a better \fscore value in $9$ data augmentation techniques in the MedSeg dataset, $16$ in MosMed, $13$ in Zenodo.
In both probabilities of $0.25$ and $0.3$, all data augmentation techniques achieved a better \fscore value in the CC-CCII dataset through the unified training strategy. The Ricord1a was the only dataset which did not achieved improvements with the unified training set. This happened due to the balancing approach that prioritized the small datasets that achieved poor results in the traditional training strategy when compared with the Ricord1a.}

Moreover, data augmentation techniques were consistently more effective with this training strategy in all \rev{six} probabilities evaluated. For a probability of $0.05$, a higher \fscore was achieved with Elastic Transform and Piecewise Affine on the CC-CCII dataset. In the MedSeg dataset, $7$ data augmentation techniques achieved a higher \fscore. In the MosMed, $12$ data augmentation techniques achieved a higher \fscore, with Piecewise Affine increasing the \fscore by $2$\%. As presented in \cref{sec:dataaugeval}, applying data augmentation in the Zenodo dataset using the standard training sets did not improve the results; only one data augmentation improved the \fscore. However, with this training strategy, $12$ data augmentation techniques achieved a higher \fscore on the Zenodo dataset. In the Ricord1a, four data augmentation techniques achieved a higher \fscore, with Piecewise Affine also increasing the \fscore by $2$\%.

For a probability of $0.1$, $5$ data augmentation techniques improved the \fscore on the CC-CCII, $7$ on the MedSeg, $13$ on the MosMed, $4$ on the Ricord1a, and $12$ on the Zenodo. Also, in MosMed, the Elastic Transform, Grid Distortion, Piecewise Affine, and Rotate increased the \fscore by $2$\%. For a probability of $0.15$, $6$ data augmentation methods improved the \fscore on the CC-CCII, $12$ on the MedSeg, $16$ on the MosMed, $9$ on the Ricord1a, and $13$ on the Zenodo. In MosMed, six data augmentation techniques increased the \fscore by $2$\%.

\rev{For a probability of $0.2$, $5$ data augmentation techniques improved the \fscore on CC-CCII, $10$ on MedSeg, $12$ on the MosMed, $8$ on the Ricord1a, and $13$ on Zenodo. In MosMed, six data augmentation techniques increased the \fscore by $2$\%, and the Shift Scale Rotate technique increased the \fscore by $3$\%.
For a probability of $0.25$, $6$ data augmentation techniques improved the \fscore on CC-CCII, $10$ on the MedSeg, $16$ on MosMed, $10$ on Ricord1a, and $15$ on Zenodo. 
In the MosMed dataset, three data augmentation techniques increased the \fscore by $3$\%.}

\rev{Finally, for a probability of $0.3$, $7$ data augmentation techniques improved the \fscore on CC-CCII, $12$ on MedSeg, $14$ on MosMed, $11$ on Ricord1a, and $13$ on  Zenodo. 
In the MosMed dataset, four data augmentation techniques increased the \fscore by $3$\%.
In general, the probabilities of $0.25$ and $0.3$ achieved the highest number of data augmentations with improvements in the \fscore values.}

\rev{The best result per dataset was: CC-CCII with Piecewise Affine applied with a probability of $0.25$, reaching an \fscore of $0.8778$; MedSeg with Grid Distortion applied with a probability of $0.3$, achieving an \fscore of $0.8994$; MosMed with Grid Distortion applied with a probability of $0.3$, attaining an \fscore of $0.8448$; Ricord1a with Gaussian Blur applied with a probability of $0.15$, achieving an \fscore of $0.8729$; Zenodo with Elastic Transform applied with a probability of $0.3$, reaching an \fscore of $0.9146$. Thus, in the CC-CCII, MedSeg, MosMed and Zenodo datasets the highest \fscore value was attained through a spatial transformation, while in Ricord1a it was achieved with a color transformation.}

\rev{With the unified training set, the spatial operations also provided the best results, confirming that these operations are the best choices for the approached problem because they encourage shape variation of the lesion regions. The Ricord1a dataset is an exception, with the Gaussian Blur achieving the highest \fscore value, suggesting that this dataset is more sensitive to color operations.}

\subsection{Experiment II: GAN-Based Data Augmentation}
\label{sec:gandataaug}

The datasets available for COVID-19 segmentation problems are limited in the number of images and have a critical imbalance problem; similar issues were identified in other medical images datasets~\citep{Banik2021imbalance,sanagavarapu2021covid}. This led to the next step of this work, which aims to develop and evaluate a domain-specific data augmentation technique based on \glspl*{gan}.

This evaluation is divided into three steps, as shown in \cref{fig:pipeline}. First, we use and evaluate two \glspl*{gan} to generate healthy \gls*{ct} images. Second, we use an encoder-decoder segmentation network to generate the lung region segmentation mask of the healthy \gls*{ct} scans generated by the \gls*{gan} models. Lastly, new lesions are added to the segmented lung regions, generating a new \gls*{ct} image with COVID-19~lesions.

\subsubsection{Healthy CT Scans Generation}
\label{experiments_gans}

The first step of the proposed data augmentation strategy is generating synthetic healthy \gls*{ct} scan samples. To this end, two well-known \glspl*{gan} were evaluated: \stargantwo~\citep{choi2020stargan} and \stylegantwoada\citep{karras2020stylegan2ada}. Regarding the \stargantwo model, it was trained for $1$M iterations, with each iteration being the processing of a batch of size $4$ through the network. Since \stargantwo is a \gls*{gan} designed to perform image-to-image translation, two domains are required to train the network. In this case, the two domains are healthy lung images and COVID-19 infected lung images. To train the \stargantwo model, the images with COVID-19 infections were taken from the Ricord1a dataset, while the healthy ones were taken from~Ricord1b. 

\begin{figure}[!htb]
\centering
\begin{center}
    \raisebox{0cm}{\textbf{\stargantwo}} \hspace{4cm} \raisebox{0cm}{\textbf{\stylegantwoada}}
\end{center}

\vspace{-6mm}

\subfloat{
\resizebox*{3.2cm}{!}{\includegraphics{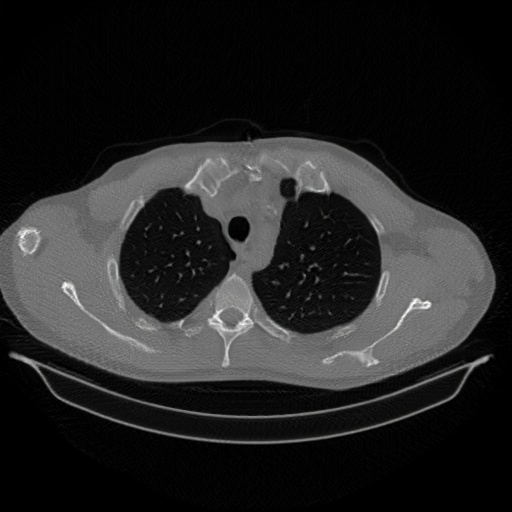}}}
\subfloat{
\resizebox*{3.2cm}{!}{\includegraphics{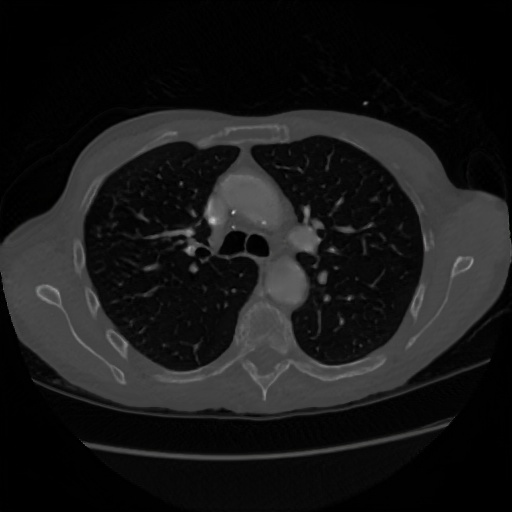}}}
\subfloat{
\hspace{.15cm}{\unskip\ \vrule\ }\hspace{.15cm}\resizebox*{3.2cm}{!}{\includegraphics{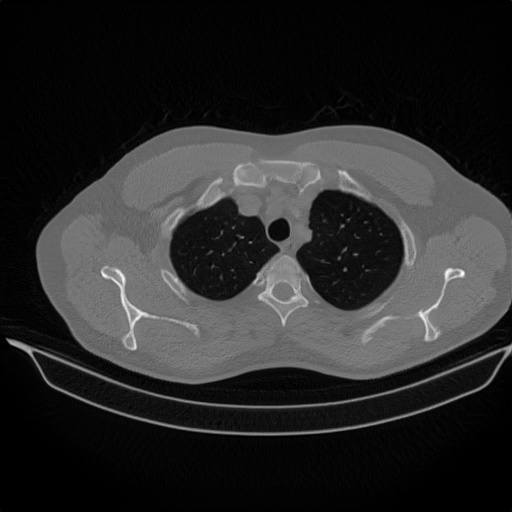}}}
\subfloat{
\resizebox*{3.2cm}{!}{\includegraphics{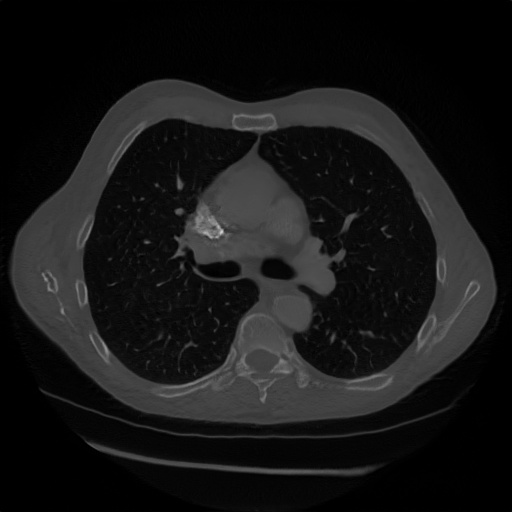}}}

\vspace{-3mm}

\subfloat{
\resizebox*{3.2cm}{!}{\includegraphics{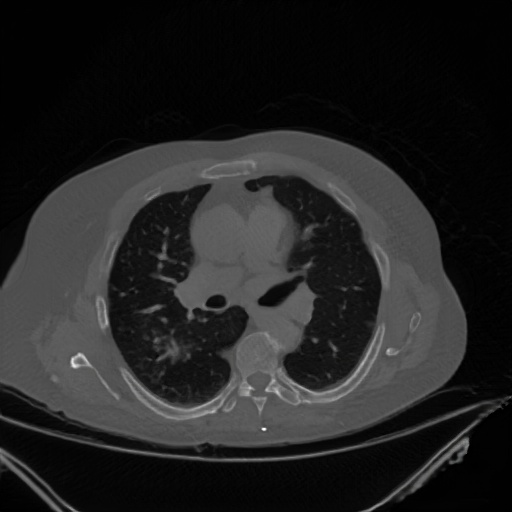}}}
\subfloat{
\resizebox*{3.2cm}{!}{\includegraphics{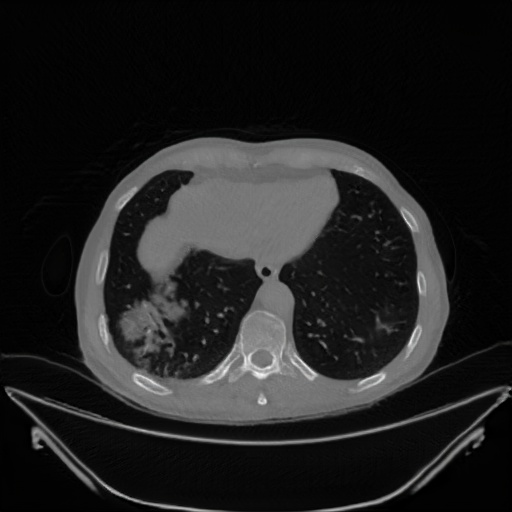}}}
\subfloat{
\hspace{.15cm}{\unskip\ \vrule\ }\hspace{.15cm}\resizebox*{3.2cm}{!}{\includegraphics{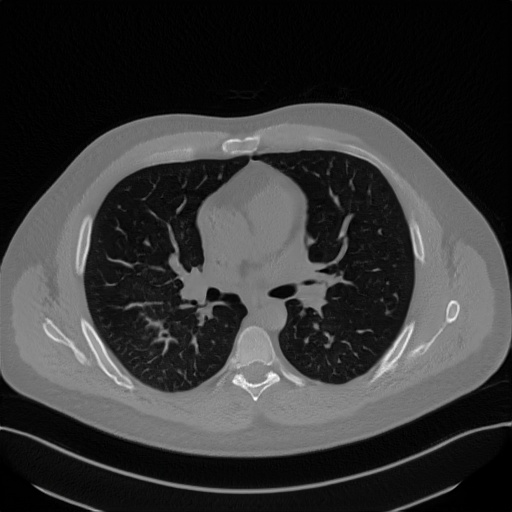}}}
\subfloat{
\resizebox*{3.2cm}{!}{\includegraphics{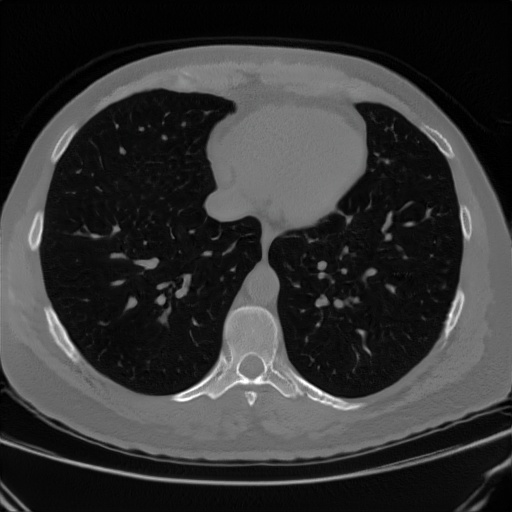}}}

\vspace{-4.5mm}

\caption[Examples of healthy lung images generated by the \stargantwo and the \stylegantwoada.]{Examples of healthy lung images generated by \stargantwo and \stylegantwoada.}
\label{fig:stargan_stylegan_generated}
\end{figure}

Regarding the \stylegantwoada model, it was trained $25$ thousand \textit{kimg}, a metric that counts the number of thousand images shown to the network~\citep{karras2020stylegan2ada}. Unlike the \stargantwo model, \stylegantwoada needed only one dataset to train the network. The dataset used in this step was the Ricord1b which provided healthy lung images. \cref{fig:stargan_stylegan_generated} presents healthy images generated with the \stargantwo (on the left) and the \stylegantwoada (on the right). The images generated with the \stylegantwoada are also promising to be used in the next steps of the data augmentation framework. As presented above, both \glspl*{gan} achieved impressive results and generated fake healthy lung images close to the real healthy lung images. However, the \stylegantwoada got a slightly better result, achieving an \gls*{fid} of 13.96, while the \stargantwo achieved an \gls*{fid} of $19.82$. In general, the \glspl*{gan} achieved promising results and generated \gls*{ct} images close to the real examples from the datasets. The next step of the proposed data augmentation will use a segmentation network to generate a segmentation mask of the lung regions of the images generated by both \glspl*{gan}. 

\subsubsection{Lung Segmentation}
\label{experiments_ls}

In the second step of the proposed data augmentation technique, a segmentation model receives the images from the \gls*{gan} models and generates the lung segmentation masks of these new images. The lung segmentation masks are explored to ensure that the new lesions are placed only within the lung region of the image. Based on the results reported in~\citep{krinski2022}, the encoder-decoder chosen in this step was the RegNetx-002 as the encoder and U-Net++ as the decoder. This combination presented impressive results in the experiments. Also, the RegNetx-002 is a relatively small network, making it faster for training and~evaluation. 

In order to train the network in this step, the CC-CCII and Zenodo datasets were combined. These datasets were the only two datasets with labels for the lung regions. This experiment also was validated through a $5$-fold cross-validation strategy, and no data augmentation was applied. 

\begin{figure}[!htb]
\centering
\begin{center}
    \raisebox{0cm}{\textbf{\stargantwo}} \hspace{4cm} \raisebox{0cm}{\textbf{\stylegantwoada}}
\end{center}

\vspace{-6mm}

\subfloat{
\resizebox*{3.2cm}{!}{\includegraphics{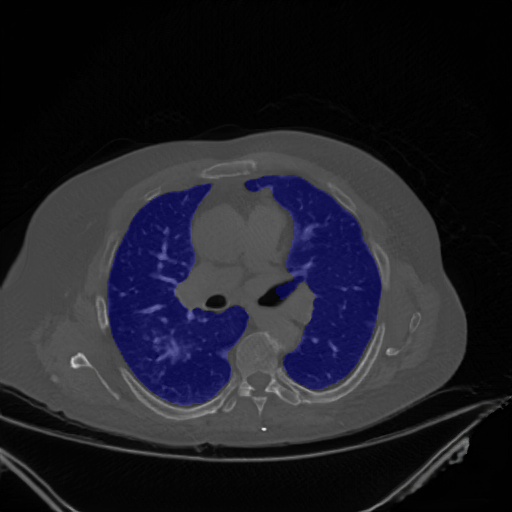}}}
\subfloat{
\resizebox*{3.2cm}{!}{\includegraphics{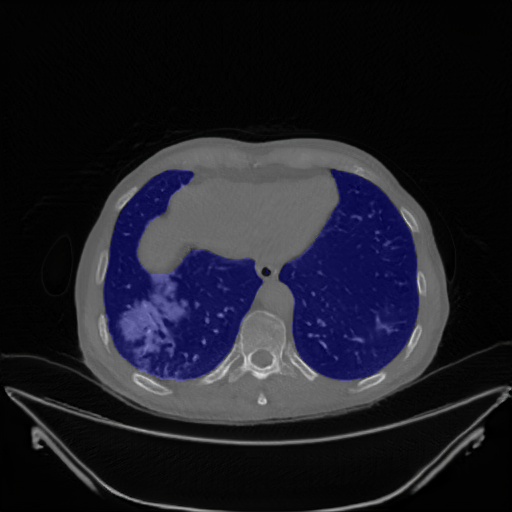}}}
\subfloat{
\hspace{.15cm}{\unskip\ \vrule\ }\hspace{.15cm}\resizebox*{3.2cm}{!}{\includegraphics{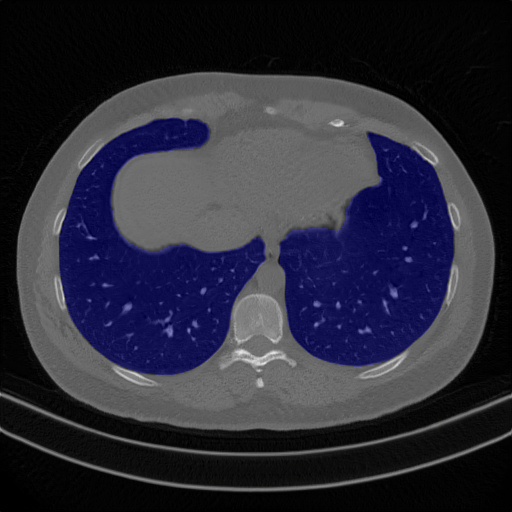}}}
\subfloat{
\resizebox*{3.2cm}{!}{\includegraphics{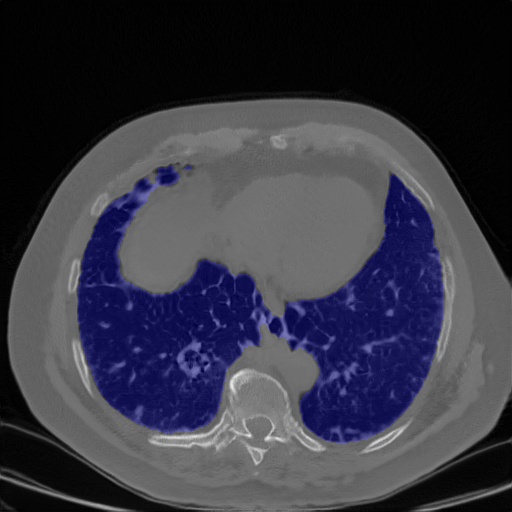}}}

\vspace{-3mm}

\subfloat{
\resizebox*{3.2cm}{!}{\includegraphics{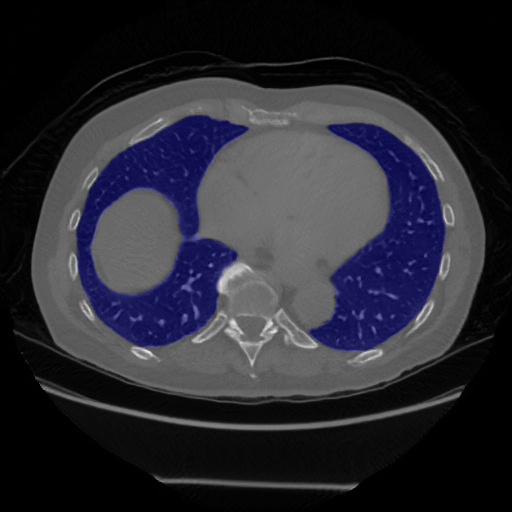}}}
\subfloat{
\resizebox*{3.2cm}{!}{\includegraphics{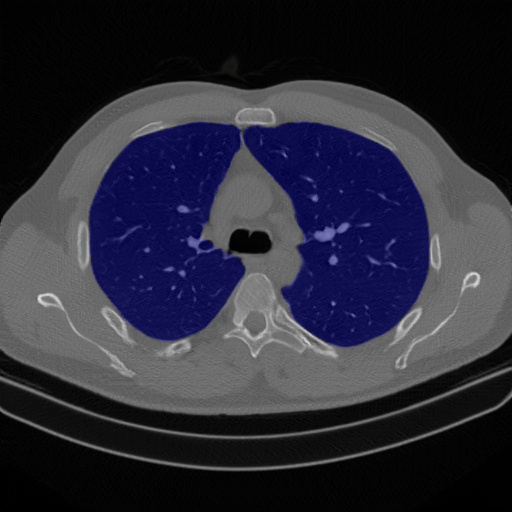}}}
\subfloat{
\hspace{.15cm}{\unskip\ \vrule\ }\hspace{.15cm}\resizebox*{3.2cm}{!}{\includegraphics{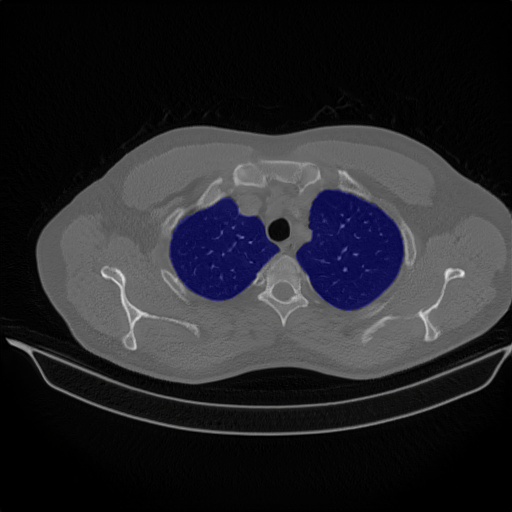}}}
\subfloat{
\resizebox*{3.2cm}{!}{\includegraphics{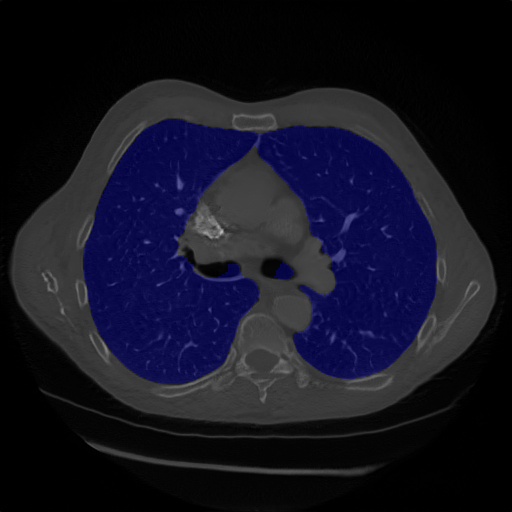}}}

\vspace{-4.5mm}

\caption[Examples of lung segmentation in the images generated by \stargantwo and \stylegantwoada.]{Examples of lung segmentation in the images generated by \stargantwo and \stylegantwoada.}
\label{fig:stargan_stylegan_seg_generated}
\end{figure}

The images generated by the \stargantwo and \stylegantwoada models were used as the test sets. That is, the segmentation network did not include images generated by the \glspl*{gan} in the training and validation steps. \cref{fig:stargan_stylegan_seg_generated} presents examples of lung segmentation in the images generated by \stargantwo~(left) and \stylegantwoada~(right). The blue-colored regions are the lung regions segmented by the network. The test sets have no ground-truth labels for quantitative analysis. However, the segmentation network achieved promising qualitative results. The generated segmentation masks are close to the expected if made through a manual labeling process, suggesting that the generated samples are visually close enough to the original~images.

\subsubsection{Adding New Lesions}
\label{experiments_anl}

The next step is the addition of COVID-19 lesions within the lung regions of the \gls*{ct} scans~(see \cref{experiments_gans}) generated by the  \glspl*{gan}~(see \cref{experiments_ls}). For each random healthy image generated by a \gls*{gan} model (\stargantwo or \stylegantwoada), a random image with lesions is chosen from the dataset Ricord1a. The first problem in this step is that the size of the lungs inside the \gls*{ct} scans varies greatly. Therefore, a matching step is performed to check if the size of the lung region of the image from the dataset is close to the size of the lung region of the healthy image generated by the \gls*{gan} model. A threshold of 10\% was used, i.e., the size of the lungs of the lesion image can be at a maximum of 10\% greater or 10\% smaller than the \gls*{gan}~image. 
\begin{figure}[!htb]
\centering
\begin{center}
    \raisebox{0cm}{\textbf{\stargantwo}} \hspace{4cm} \raisebox{0cm}{\textbf{\stylegantwoada}}
\end{center}

\vspace{-6mm}

\subfloat{
\resizebox*{3.2cm}{!}{\includegraphics{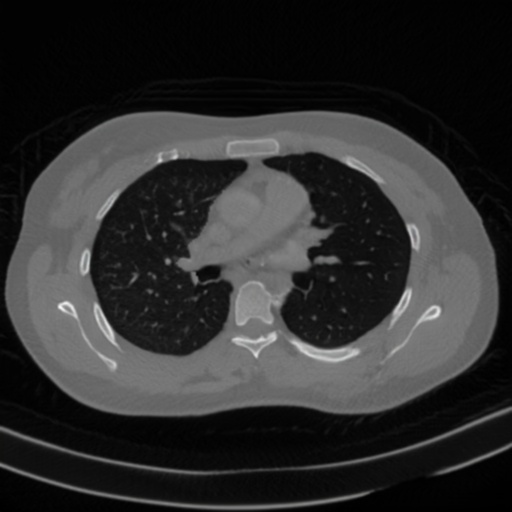}}}
\subfloat{
\resizebox*{3.2cm}{!}{\includegraphics{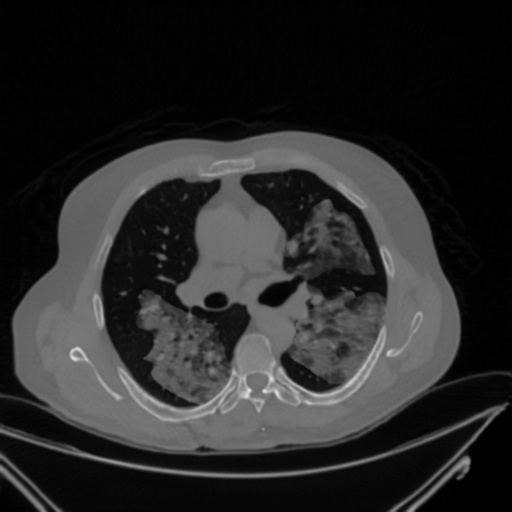}}}
\subfloat{
\hspace{.15cm}{\unskip\ \vrule\ }\hspace{.15cm}\resizebox*{3.2cm}{!}{\includegraphics{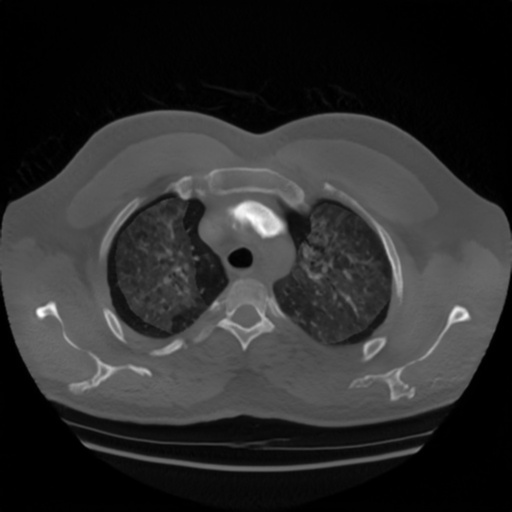}}}
\subfloat{
\resizebox*{3.2cm}{!}{\includegraphics{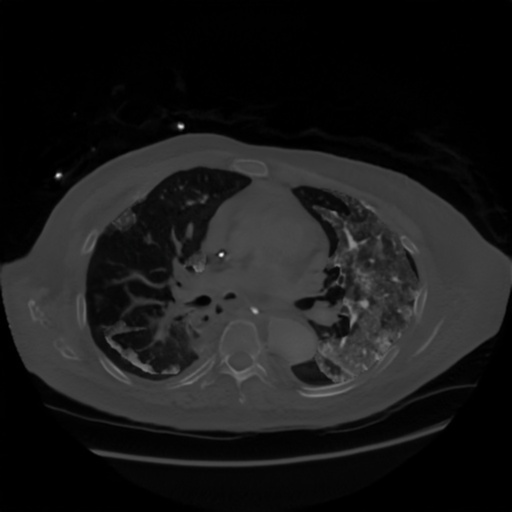}}}

\vspace{-3mm}

\subfloat{
\resizebox*{3.2cm}{!}{\includegraphics{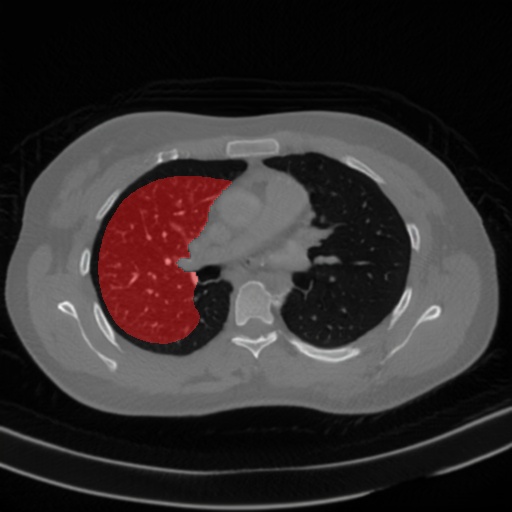}}}
\subfloat{
\resizebox*{3.2cm}{!}{\includegraphics{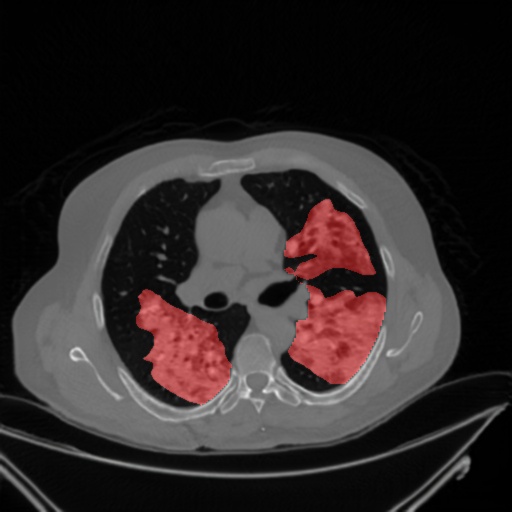}}}
\subfloat{
\hspace{.15cm}{\unskip\ \vrule\ }\hspace{.15cm}\resizebox*{3.2cm}{!}{\includegraphics{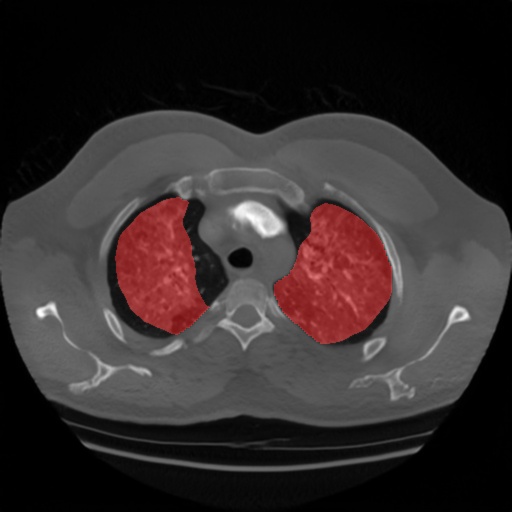}}}
\subfloat{
\resizebox*{3.2cm}{!}{\includegraphics{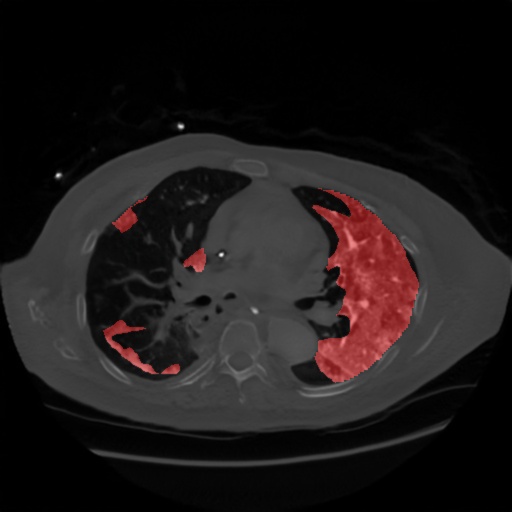}}}

\vspace{-4mm}

\caption[Examples of images generated by the Stargan and the Stylegan after the addition of COVID-19 lesions.]{Examples of images generated by \stargantwo and \stylegantwoada after adding COVID-19 lesions and the corresponding generated segmentation masks.}
\label{fig:stargan_stylegan_nl1}
\end{figure}

After matching the lesion image from the dataset with the image generated by the \gls*{gan} model, the lungs of the dataset image are cropped from the original image and positioned on top of the \gls*{gan} image. The center of the bounding box of the lungs is calculated and used to align the lungs of the dataset image with the lungs of the \gls*{gan} image. Then, the functions AddWeighted and GaussianBlur from the OpenCV Library~\citep{opencv_library} are applied to mix the images and smooth out the edges, respectively. \cref{fig:stargan_stylegan_nl1} illustrates the resulting images and the corresponding segmentation masks with lesions added to images generated by \stargantwo~(left) and \stylegantwoada~(right). As expected, due to the promising lung segmentation step presented in \cref{experiments_ls}, the COVID-19 lesions were correctly placed within the lung~regions. 

\subsubsection{Proposed GAN-Based Data Augmentation Evaluation}
\label{sec:ganeval}

\cref{resultsgan} \rev{(a continuation of Tables~\ref{results3},~\ref{results4} and \ref{results6})} presents the evaluation results for the proposed data augmentation technique in the unified training set. It was evaluated with two variations (with and without flip). First, the lesions from the dataset were added to the healthy lung images generated by the \glspl*{gan} without altering the lesion images. Then, in the second variation, the lesion images were horizontally flipped before being added to the healthy image generated by the \glspl*{gan}.

\begin{table}[!ht]
\centering
\tbl{Results of the data augmentation evaluation when unifying the training sets (Continuation of Tables~\ref{results3}, \ref{results4} \rev{and \ref{results6}}. $p$ stands for probability, the blue-colored values indicate the best \fscore values, and the red-colored values indicate the best \gls*{iou} values. The values highlighted in green show the data augmentation techniques in which the P-value achieved values lower than $0.05$, and thus the null hypothesis was rejected (i.e., there is a statistical difference and the results achieved are better than without data augmentation). Two variations of the proposed algorithm were evaluated. The + F indicates the variation in which the lesion images were horizontally flipped before being added to the healthy image generated by the~\glspl*{gan}.}
{\resizebox*{\textwidth}{!}{\begin{tabular}{cccccccccccc} \toprule
\multicolumn{1}{c}{\textbf{$p$}} &
\multicolumn{1}{c}{\textbf{Augmentation}} &
\multicolumn{2}{c}{\textbf{CC-CCII}} &
\multicolumn{2}{c}{\textbf{MedSeg}} &
\multicolumn{2}{c}{\textbf{MosMed}} &
\multicolumn{2}{c}{\textbf{Ricord1a}} &
\multicolumn{2}{c}{\textbf{Zenodo}} \\ \midrule
\makecell{} &
\makecell{} &
\makecell{\fscore} &
\makecell{\gls*{iou}} & 
\makecell{\fscore} &
\makecell{\gls*{iou}} & 
\makecell{\fscore} &
\makecell{\gls*{iou}} & 
\makecell{\fscore} &
\makecell{\gls*{iou}} & 
\makecell{\fscore} &
\makecell{\gls*{iou}} \\ \midrule
\makecell{} &
\makecell{No Augmentation}&
\makecell{0.8636}&
\makecell{0.8087}&
\makecell{0.8881}&
\makecell{0.8253}&
\makecell{0.8185}&
\makecell{0.7547}&
\makecell{0.8599}&
\makecell{0.7947}&
\makecell{0.9096}&
\makecell{0.8514} \\ \midrule
\makecell{0.05} & 
\makecell{\colorbox{white}{\makebox(21,2){\stargantwo}}\\
\colorbox{white}{\makebox(21,2){\stargantwo+ F}}\\
\colorbox{white}{\makebox(21,2){\stylegantwoada}}\\
\colorbox{white}{\makebox(21,2){\stylegantwoada+ F}}\\}&
\makecell{\colorbox{white}{\makebox(21,2){0.8615}}\\
\colorbox{white}{\makebox(21,2){0.8628}}\\
\colorbox{white}{\makebox(21,2){0.8624}}\\
\colorbox{white}{\makebox(21,2){0.8644}}}&
\makecell{\colorbox{white}{\makebox(21,2){0.8060}}\\
\colorbox{white}{\makebox(21,2){0.8071}}\\
\colorbox{white}{\makebox(21,2){0.8071}}\\
\colorbox{white}{\makebox(21,2){0.8085}}}&
\makecell{\colorbox{white}{\makebox(21,2){0.8893}}\\
\colorbox{white}{\makebox(21,2){0.8873}}\\
\colorbox{white}{\makebox(21,2){0.8861}}\\
\colorbox{white}{\makebox(21,2){0.8873}}}&
\makecell{\colorbox{white}{\makebox(21,2){0.8270}}\\
\colorbox{white}{\makebox(21,2){0.8247}}\\
\colorbox{white}{\makebox(21,2){0.8235}}\\
\colorbox{white}{\makebox(21,2){0.8246}}}&
\makecell{\colorbox{green!30}{\makebox(21,2){0.8200}}\\
\colorbox{white}{\makebox(21,2){0.8193}}\\
\colorbox{white}{\makebox(21,2){0.8180}}\\
\colorbox{white}{\makebox(21,2){0.8163}}}&
\makecell{\colorbox{white}{\makebox(21,2){0.7565}}\\
\colorbox{white}{\makebox(21,2){0.7566}}\\
\colorbox{white}{\makebox(21,2){0.7546}}\\
\colorbox{white}{\makebox(21,2){0.7534}}}&
\makecell{\colorbox{green!30}{\makebox(21,2){0.8653}}\\
\colorbox{green!30}{\makebox(21,2){0.8620}}\\
\colorbox{white}{\makebox(21,2){0.8535}}\\
\colorbox{white}{\makebox(21,2){0.8592}}}&
\makecell{\colorbox{white}{\makebox(21,2){0.8009}}\\
\colorbox{white}{\makebox(21,2){0.7974}}\\
\colorbox{white}{\makebox(21,2){0.7872}}\\
\colorbox{white}{\makebox(21,2){0.7938}}}&
\makecell{\colorbox{green!30}{\makebox(21,2){0.9107}}\\
\colorbox{green!30}{\makebox(21,2){0.9106}}\\
\colorbox{white}{\makebox(21,2){0.9076}}\\
\colorbox{green!30}{\makebox(21,2){0.9098}}}&
\makecell{\colorbox{white}{\makebox(21,2){0.8533}}\\
\colorbox{white}{\makebox(21,2){0.8528}}\\
\colorbox{white}{\makebox(21,2){0.8491}}\\
\colorbox{white}{\makebox(21,2){0.8518}}}\\ \midrule
\makecell{0.1} & 
\makecell{\colorbox{white}{\makebox(21,2){\stargantwo}}\\
\colorbox{white}{\makebox(21,2){\stargantwo + F}}\\
\colorbox{white}{\makebox(21,2){\stylegantwoada}}\\
\colorbox{white}{\makebox(21,2){\stylegantwoada + F}}\\}&
\makecell{\colorbox{white}{\makebox(21,2){0.8632}}\\
\colorbox{white}{\makebox(21,2){0.8623}}\\
\colorbox{white}{\makebox(21,2){0.8617}}\\
\colorbox{white}{\makebox(21,2){0.8616}}}&
\makecell{\colorbox{white}{\makebox(21,2){0.8082}}\\
\colorbox{white}{\makebox(21,2){0.8058}}\\
\colorbox{white}{\makebox(21,2){0.8061}}\\
\colorbox{white}{\makebox(21,2){0.8058}}}&
\makecell{\colorbox{white}{\makebox(21,2){0.8853}}\\
\colorbox{white}{\makebox(21,2){0.8871}}\\
\colorbox{white}{\makebox(21,2){0.8879}}\\
\colorbox{white}{\makebox(21,2){0.8891}}}&
\makecell{\colorbox{white}{\makebox(21,2){0.8222}}\\
\colorbox{white}{\makebox(21,2){0.8244}}\\
\colorbox{white}{\makebox(21,2){0.8251}}\\
\colorbox{white}{\makebox(21,2){0.8268}}}&
\makecell{\colorbox{white}{\makebox(21,2){0.8133}}\\
\colorbox{white}{\makebox(21,2){0.8176}}\\
\colorbox{green!30}{\makebox(21,2){0.8213}}\\
\colorbox{green!30}{\makebox(21,2){0.8227}}}&
\makecell{\colorbox{white}{\makebox(21,2){0.7507}}\\
\colorbox{white}{\makebox(21,2){0.7545}}\\
\colorbox{white}{\makebox(21,2){0.7585}}\\
\colorbox{white}{\makebox(21,2){0.7598}}}&
\makecell{\colorbox{white}{\makebox(21,2){0.8560}}\\
\colorbox{green!30}{\makebox(21,2){0.8627}}\\
\colorbox{green!30}{\makebox(21,2){0.8672}}\\
\colorbox{green!30}{\makebox(21,2){\textbf{\textcolor{blue}{0.8711}}}}}&
\makecell{\colorbox{white}{\makebox(21,2){0.7908}}\\
\colorbox{white}{\makebox(21,2){0.7980}}\\
\colorbox{white}{\makebox(21,2){0.8034}}\\
\colorbox{white}{\makebox(21,2){\textbf{\textcolor{red}{0.8081}}}}}&
\makecell{\colorbox{white}{\makebox(21,2){0.9082}}\\
\colorbox{white}{\makebox(21,2){0.9091}}\\
\colorbox{green!30}{\makebox(21,2){0.9117}}\\
\colorbox{green!30}{\makebox(21,2){0.9102}}}&
\makecell{\colorbox{white}{\makebox(21,2){0.8499}}\\
\colorbox{white}{\makebox(21,2){0.8508}}\\
\colorbox{white}{\makebox(21,2){0.8540}}\\
\colorbox{white}{\makebox(21,2){0.8527}}} \\ \midrule
\makecell{0.15} &   
\makecell{\colorbox{white}{\makebox(21,2){\stargantwo}}\\
\colorbox{white}{\makebox(21,2){\stargantwo + F}}\\
\colorbox{white}{\makebox(21,2){\stylegantwoada}}\\
\colorbox{white}{\makebox(21,2){\stylegantwoada + F}}\\}&
\makecell{\colorbox{white}{\makebox(21,2){0.8637}}\\
\colorbox{white}{\makebox(21,2){0.8654}}\\
\colorbox{white}{\makebox(21,2){0.8645}}\\
\colorbox{white}{\makebox(21,2){0.8622}}}&
\makecell{\colorbox{white}{\makebox(21,2){0.8087}}\\
\colorbox{white}{\makebox(21,2){0.8096}}\\
\colorbox{white}{\makebox(21,2){0.8086}}\\
\colorbox{white}{\makebox(21,2){0.8066}}}&
\makecell{\colorbox{white}{\makebox(21,2){0.8870}}\\
\colorbox{white}{\makebox(21,2){0.8871}}\\
\colorbox{white}{\makebox(21,2){0.8872}}\\
\colorbox{white}{\makebox(21,2){0.8885}}}&
\makecell{\colorbox{white}{\makebox(21,2){0.8243}}\\
\colorbox{white}{\makebox(21,2){0.8248}}\\
\colorbox{white}{\makebox(21,2){0.8239}}\\
\colorbox{white}{\makebox(21,2){0.8260}}}&
\makecell{\colorbox{green!30}{\makebox(21,2){0.8214}}\\
\colorbox{green!30}{\makebox(21,2){0.8239}}\\
\colorbox{white}{\makebox(21,2){0.8155}}\\
\colorbox{green!30}{\makebox(21,2){0.8215}}}&
\makecell{\colorbox{white}{\makebox(21,2){0.7586}}\\
\colorbox{white}{\makebox(21,2){0.7610}}\\
\colorbox{white}{\makebox(21,2){0.7528}}\\
\colorbox{white}{\makebox(21,2){0.7586}}}&
\makecell{\colorbox{green!30}{\makebox(21,2){0.8689}}\\
\colorbox{green!30}{\makebox(21,2){0.8665}}\\
\colorbox{green!30}{\makebox(21,2){0.8612}}\\
\colorbox{green!30}{\makebox(21,2){0.8696}}}&
\makecell{\colorbox{white}{\makebox(21,2){0.8055}}\\
\colorbox{white}{\makebox(21,2){0.8031}}\\
\colorbox{white}{\makebox(21,2){0.7965}}\\
\colorbox{white}{\makebox(21,2){0.8062}}}&
\makecell{\colorbox{white}{\makebox(21,2){0.9099}}\\
\colorbox{green!30}{\makebox(21,2){0.9107}}\\
\colorbox{white}{\makebox(21,2){0.9090}}\\
\colorbox{green!30}{\makebox(21,2){0.9115}}}&
\makecell{\colorbox{white}{\makebox(21,2){0.8521}}\\
\colorbox{white}{\makebox(21,2){0.8527}}\\
\colorbox{white}{\makebox(21,2){0.8509}}\\
\colorbox{white}{\makebox(21,2){0.8541}}} \\ \midrule
\makecell{0.2} &
\makecell{\colorbox{white}{\makebox(21,2){\stargantwo}}\\
\colorbox{white}{\makebox(21,2){\stargantwo + F}}\\
\colorbox{white}{\makebox(21,2){\stylegantwoada}}\\
\colorbox{white}{\makebox(21,2){\stylegantwoada + F}}\\}&
\makecell{\colorbox{white}{\makebox(21,2){0.8644}}\\
\colorbox{white}{\makebox(21,2){0.8655}}\\
\colorbox{white}{\makebox(21,2){0.8635}}\\
\colorbox{white}{\makebox(21,2){0.8614}}}&
\makecell{\colorbox{white}{\makebox(21,2){0.8085}}\\
\colorbox{white}{\makebox(21,2){0.8098}}\\
\colorbox{white}{\makebox(21,2){0.8093}}\\
\colorbox{white}{\makebox(21,2){0.8065}}}&
\makecell{\colorbox{white}{\makebox(21,2){0.8860}}\\
\colorbox{white}{\makebox(21,2){0.8867}}\\
\colorbox{white}{\makebox(21,2){0.8886}}\\
\colorbox{white}{\makebox(21,2){0.8895}}}&
\makecell{\colorbox{white}{\makebox(21,2){0.8231}}\\
\colorbox{white}{\makebox(21,2){0.8242}}\\
\colorbox{white}{\makebox(21,2){0.8256}}\\
\colorbox{white}{\makebox(21,2){0.8269}}}&
\makecell{\colorbox{white}{\makebox(21,2){0.8201}}\\
\colorbox{white}{\makebox(21,2){0.8192}}\\
\colorbox{white}{\makebox(21,2){0.8190}}\\
\colorbox{white}{\makebox(21,2){0.8201}}}&
\makecell{\colorbox{white}{\makebox(21,2){0.7563}}\\
\colorbox{white}{\makebox(21,2){0.7556}}\\
\colorbox{white}{\makebox(21,2){0.7558}}\\
\colorbox{white}{\makebox(21,2){0.7571}}}&
\makecell{\colorbox{white}{\makebox(21,2){0.8555}}\\
\colorbox{green!30}{\makebox(21,2){0.8626}}\\
\colorbox{white}{\makebox(21,2){0.8607}}\\
\colorbox{green!30}{\makebox(21,2){\textbf{\textcolor{blue}{0.8729}}}}}&
\makecell{\colorbox{white}{\makebox(21,2){0.7901}}\\
\colorbox{white}{\makebox(21,2){0.7983}}\\
\colorbox{white}{\makebox(21,2){0.7959}}\\
\colorbox{white}{\makebox(21,2){\textbf{\textcolor{red}{0.8103}}}}}&
\makecell{\colorbox{white}{\makebox(21,2){0.9084}}\\
\colorbox{white}{\makebox(21,2){0.9097}}\\
\colorbox{white}{\makebox(21,2){0.9103}}\\
\colorbox{white}{\makebox(21,2){0.9104}}}&
\makecell{\colorbox{white}{\makebox(21,2){0.8499}}\\
\colorbox{white}{\makebox(21,2){0.8518}}\\
\colorbox{white}{\makebox(21,2){0.8522}}\\
\colorbox{green!30}{\makebox(21,2){0.8528}}} \\ \midrule
\makecell{0.25} & 
\makecell{\colorbox{white}{\makebox(21,2){\stargantwo}}\\
\colorbox{white}{\makebox(21,2){\stargantwo+ F}}\\
\colorbox{white}{\makebox(21,2){\stylegantwoada}}\\
\colorbox{white}{\makebox(21,2){\stylegantwoada+ F}}\\}&
\makecell{\colorbox{white}{\makebox(21,2){0.8660}}\\
\colorbox{white}{\makebox(21,2){0.8643}}\\
\colorbox{white}{\makebox(21,2){0.8617}}\\
\colorbox{white}{\makebox(21,2){0.8668}}}&
\makecell{\colorbox{white}{\makebox(21,2){0.8109}}\\
\colorbox{white}{\makebox(21,2){0.8081}}\\
\colorbox{white}{\makebox(21,2){0.8060}}\\
\colorbox{white}{\makebox(21,2){0.8116}}}&
\makecell{\colorbox{white}{\makebox(21,2){0.8891}}\\
\colorbox{white}{\makebox(21,2){0.8893}}\\
\colorbox{white}{\makebox(21,2){0.8878}}\\
\colorbox{white}{\makebox(21,2){0.8875}}}&
\makecell{\colorbox{white}{\makebox(21,2){0.8264}}\\
\colorbox{white}{\makebox(21,2){0.8262}}\\
\colorbox{white}{\makebox(21,2){0.8251}}\\
\colorbox{white}{\makebox(21,2){0.8245}}}&
\makecell{\colorbox{white}{\makebox(21,2){0.8189}}\\
\colorbox{white}{\makebox(21,2){0.8192}}\\
\colorbox{white}{\makebox(21,2){0.8183}}\\
\colorbox{green!30}{\makebox(21,2){0.8222}}}&
\makecell{\colorbox{white}{\makebox(21,2){0.7562}}\\
\colorbox{white}{\makebox(21,2){0.7558}}\\
\colorbox{white}{\makebox(21,2){0.7548}}\\
\colorbox{white}{\makebox(21,2){0.7583}}}&
\makecell{\colorbox{green!30}{\makebox(21,2){0.8664}}\\
\colorbox{green!30}{\makebox(21,2){0.8693}}\\
\colorbox{green!30}{\makebox(21,2){0.8619}}\\
\colorbox{green!30}{\makebox(21,2){0.8656}}}&
\makecell{\colorbox{white}{\makebox(21,2){0.8029}}\\
\colorbox{white}{\makebox(21,2){0.8061}}\\
\colorbox{white}{\makebox(21,2){0.7976}}\\
\colorbox{white}{\makebox(21,2){0.8017}}}&
\makecell{\colorbox{white}{\makebox(21,2){0.9097}}\\
\colorbox{white}{\makebox(21,2){0.9090}}\\
\colorbox{white}{\makebox(21,2){0.9101}}\\
\colorbox{white}{\makebox(21,2){0.9099}}}&
\makecell{\colorbox{white}{\makebox(21,2){0.8518}}\\
\colorbox{white}{\makebox(21,2){0.8510}}\\
\colorbox{white}{\makebox(21,2){0.8520}}\\
\colorbox{white}{\makebox(21,2){0.8520}}} \\ \midrule
\makecell{0.3} & 
\makecell{\colorbox{white}{\makebox(21,2){\stargantwo}}\\
\colorbox{white}{\makebox(21,2){\stargantwo+ F}}\\
\colorbox{white}{\makebox(21,2){\stylegantwoada}}\\
\colorbox{white}{\makebox(21,2){\stylegantwoada+ F}}\\}&
\makecell{\colorbox{white}{\makebox(21,2){0.8635}}\\
\colorbox{white}{\makebox(21,2){0.8645}}\\
\colorbox{white}{\makebox(21,2){0.8648}}\\
\colorbox{green!30}{\makebox(21,2){0.8675}}}&
\makecell{\colorbox{white}{\makebox(21,2){0.8078}}\\
\colorbox{white}{\makebox(21,2){0.8087}}\\
\colorbox{white}{\makebox(21,2){0.8100}}\\
\colorbox{white}{\makebox(21,2){0.8119}}}&
\makecell{\colorbox{white}{\makebox(21,2){0.8851}}\\
\colorbox{white}{\makebox(21,2){0.8884}}\\
\colorbox{white}{\makebox(21,2){0.8877}}\\
\colorbox{white}{\makebox(21,2){0.8876}}}&
\makecell{\colorbox{white}{\makebox(21,2){0.8221}}\\
\colorbox{white}{\makebox(21,2){0.8259}}\\
\colorbox{white}{\makebox(21,2){0.8247}}\\
\colorbox{white}{\makebox(21,2){0.8247}}}&
\makecell{\colorbox{white}{\makebox(21,2){0.8204}}\\
\colorbox{green!30}{\makebox(21,2){0.8214}}\\
\colorbox{white}{\makebox(21,2){0.8197}}\\
\colorbox{green!30}{\makebox(21,2){0.8195}}}&
\makecell{\colorbox{white}{\makebox(21,2){0.7572}}\\
\colorbox{white}{\makebox(21,2){0.7584}}\\
\colorbox{white}{\makebox(21,2){0.7562}}\\
\colorbox{white}{\makebox(21,2){0.7568}}}&
\makecell{\colorbox{green!30}{\makebox(21,2){0.8617}}\\
\colorbox{green!30}{\makebox(21,2){\textbf{\textcolor{blue}{0.8737}}}}\\
\colorbox{green!30}{\makebox(21,2){0.8620}}\\
\colorbox{green!30}{\makebox(21,2){0.8721}}}&
\makecell{\colorbox{white}{\makebox(21,2){0.7975}}\\
\colorbox{white}{\makebox(21,2){\textbf{\textcolor{red}{0.8111}}}}\\
\colorbox{white}{\makebox(21,2){0.7975}}\\
\colorbox{white}{\makebox(21,2){0.8095}}}&
\makecell{\colorbox{white}{\makebox(21,2){0.9088}}\\
\colorbox{green!30}{\makebox(21,2){0.9105}}\\
\colorbox{white}{\makebox(21,2){0.9095}}\\
\colorbox{green!30}{\makebox(21,2){0.9105}}}&
\makecell{\colorbox{white}{\makebox(21,2){0.8504}}\\
\colorbox{white}{\makebox(21,2){0.8532}}\\
\colorbox{white}{\makebox(21,2){0.8512}}\\
\colorbox{white}{\makebox(21,2){0.8530}}}\\ \bottomrule
\end{tabular}}}
\label{resultsgan}
\end{table}

\begin{figure}[!htb]
\centering
\captionsetup[subfigure]{labelformat=empty}

\resizebox{0.95\linewidth}{!}{
\subfloat{
\resizebox*{3.5cm}{!}{\includegraphics{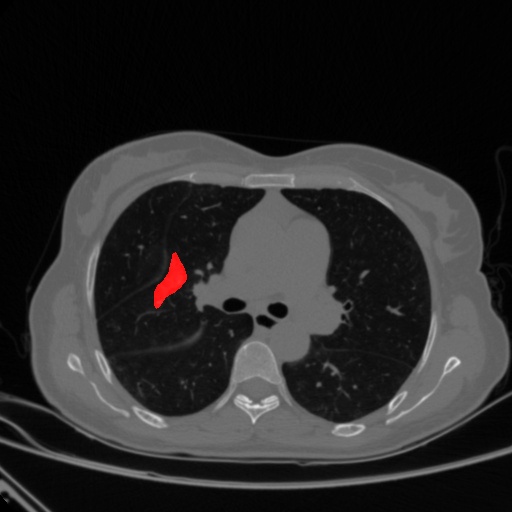}}}
\subfloat{
\resizebox*{3.5cm}{!}{\includegraphics{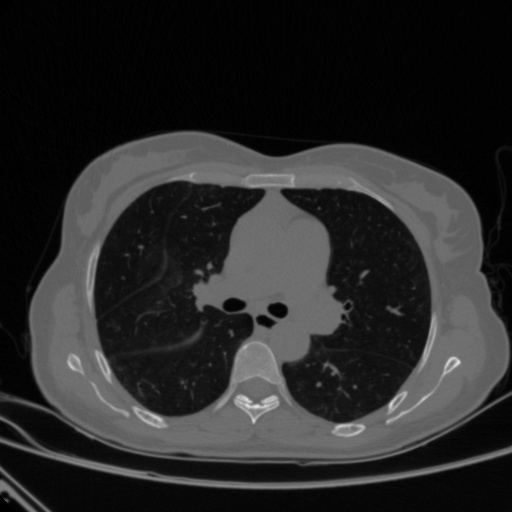}}}
\subfloat{
\resizebox*{3.5cm}{!}{\includegraphics{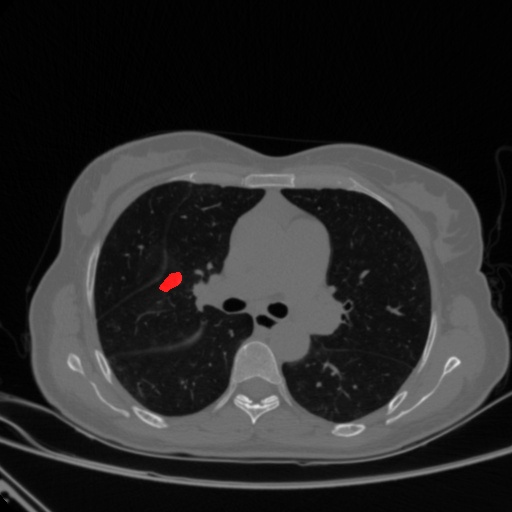}}}
\subfloat{
\resizebox*{3.5cm}{!}{\includegraphics{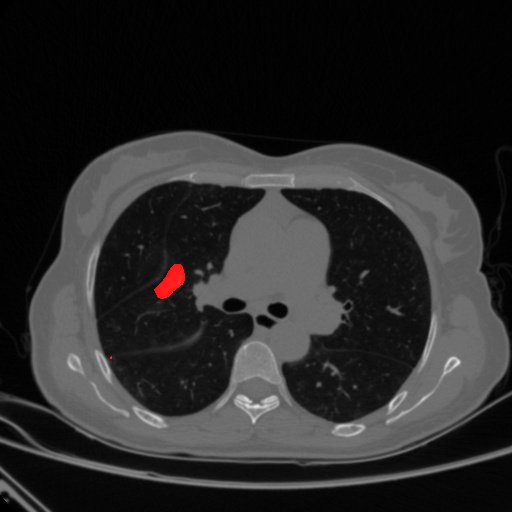}}}
\subfloat{
\resizebox*{3.5cm}{!}{\includegraphics{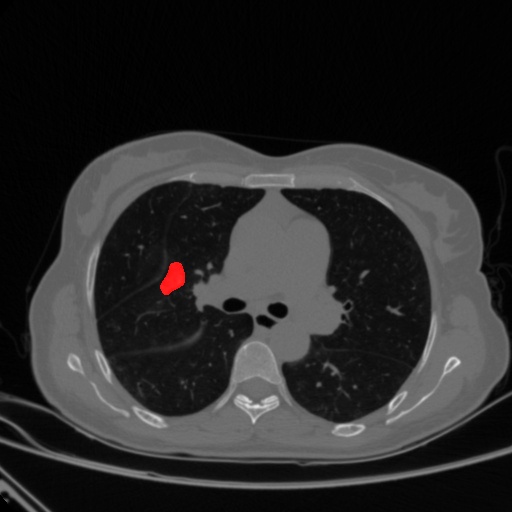}}}
\subfloat{
\resizebox*{3.5cm}{!}{\includegraphics{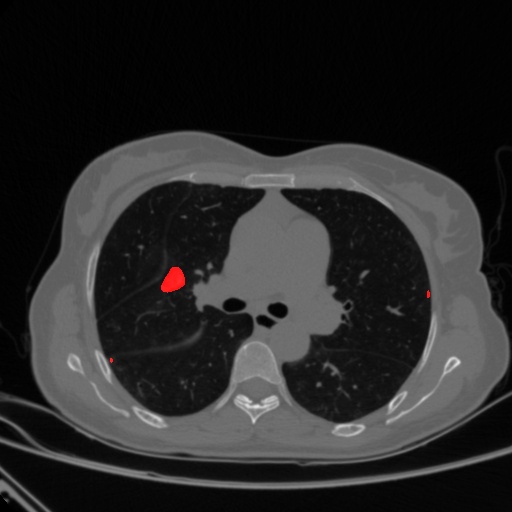}}}
}

\vspace{-1.85mm}

\resizebox{0.95\linewidth}{!}{
\subfloat{
\resizebox*{3.5cm}{!}{\includegraphics{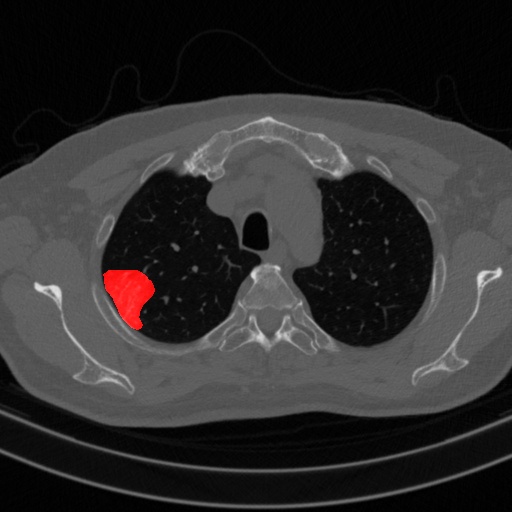}}}
\subfloat{
\resizebox*{3.5cm}{!}{\includegraphics{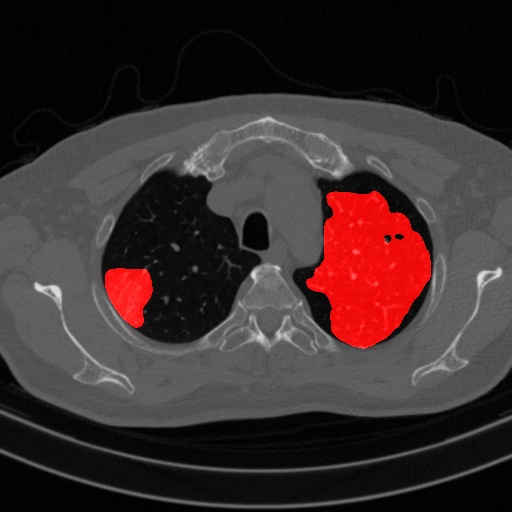}}}
\subfloat{
\resizebox*{3.5cm}{!}{\includegraphics{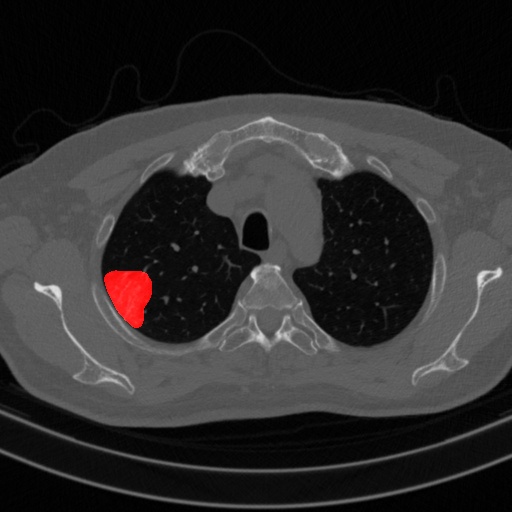}}}
\subfloat{
\resizebox*{3.5cm}{!}{\includegraphics{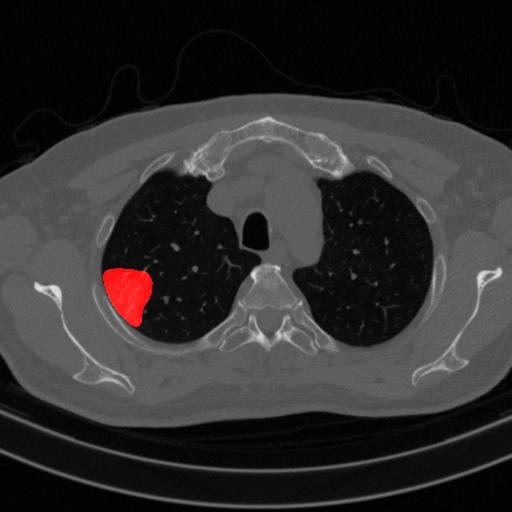}}}
\subfloat{
\resizebox*{3.5cm}{!}{\includegraphics{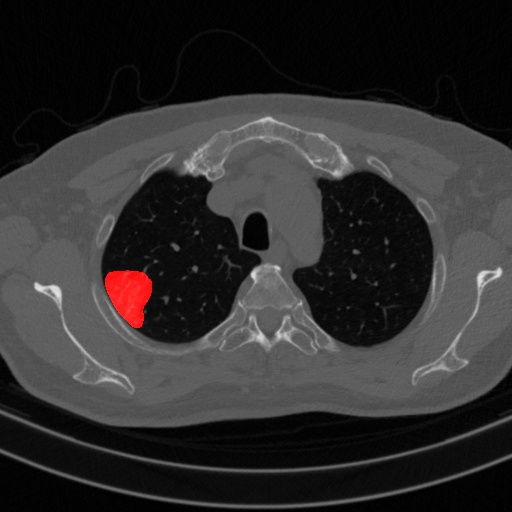}}}
\subfloat{
\resizebox*{3.5cm}{!}{\includegraphics{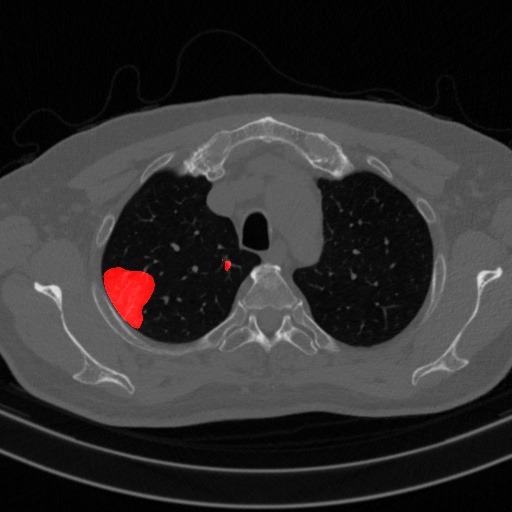}}} 
}

\vspace{-1.85mm}

\resizebox{0.95\linewidth}{!}{
\subfloat{
\resizebox*{3.5cm}{!}{\includegraphics{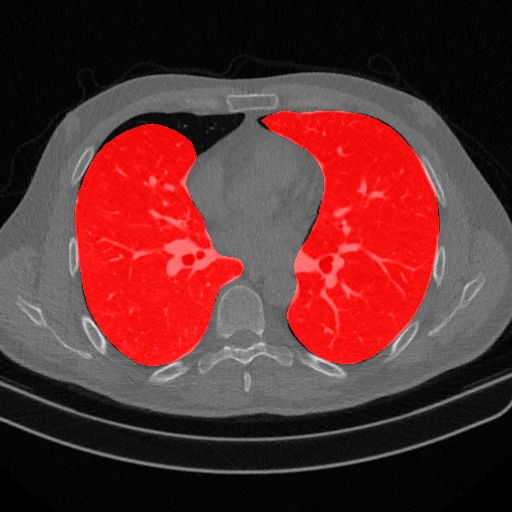}}}
\subfloat{
\resizebox*{3.5cm}{!}{\includegraphics{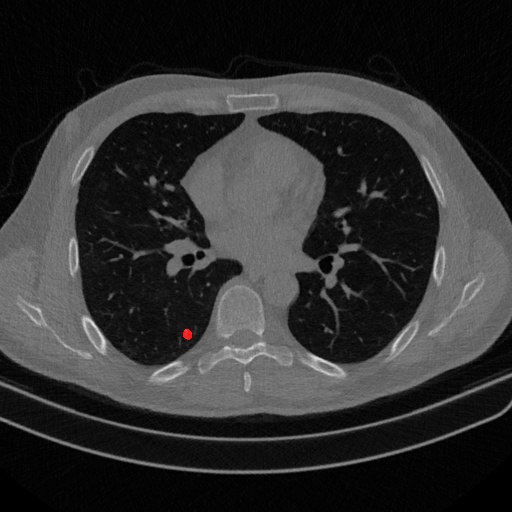}}}
\subfloat{
\resizebox*{3.5cm}{!}{\includegraphics{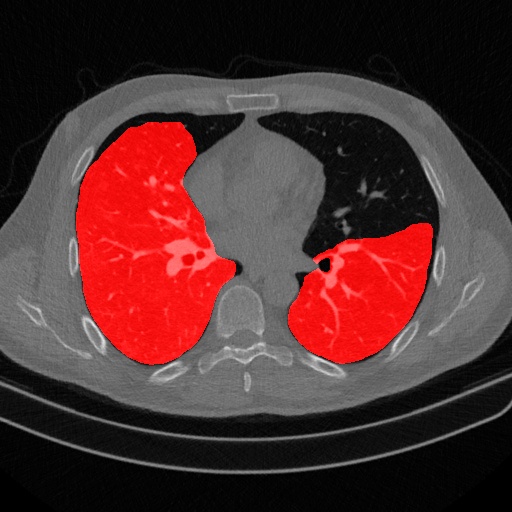}}}
\subfloat{
\resizebox*{3.5cm}{!}{\includegraphics{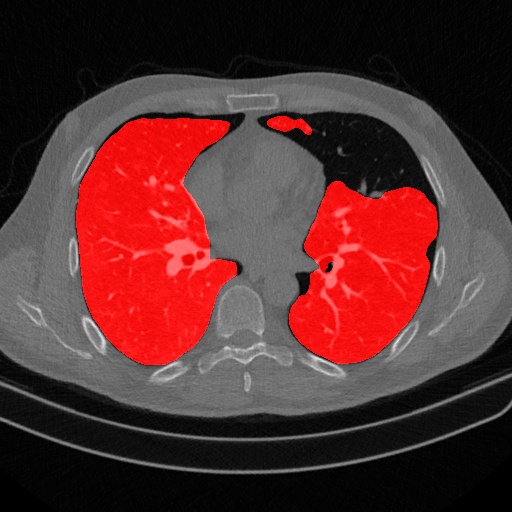}}}
\subfloat{
\resizebox*{3.5cm}{!}{\includegraphics{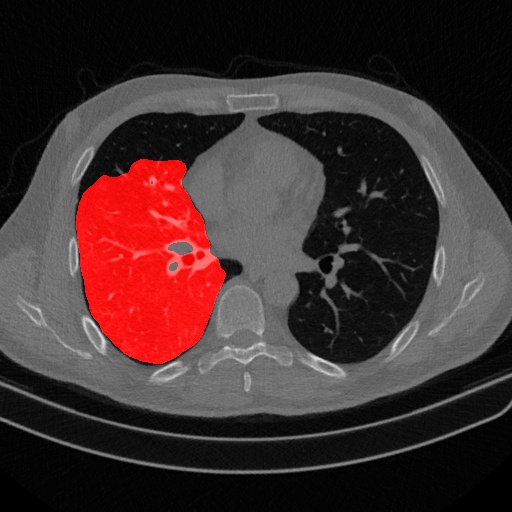}}}
\subfloat{
\resizebox*{3.5cm}{!}{\includegraphics{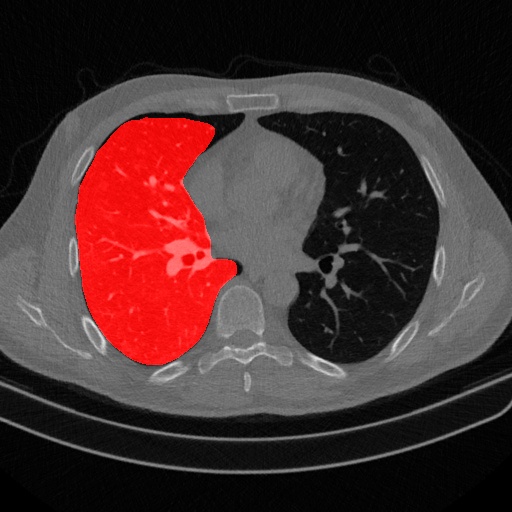}}}
}

\vspace{-1.85mm}

\resizebox{0.95\linewidth}{!}{
\subfloat[GT]{
\resizebox*{3.5cm}{!}{\includegraphics{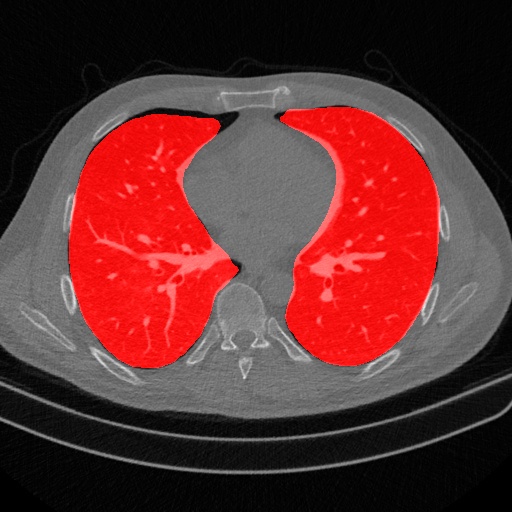}}}
\subfloat[No Data Aug]{
\resizebox*{3.5cm}{!}{\includegraphics{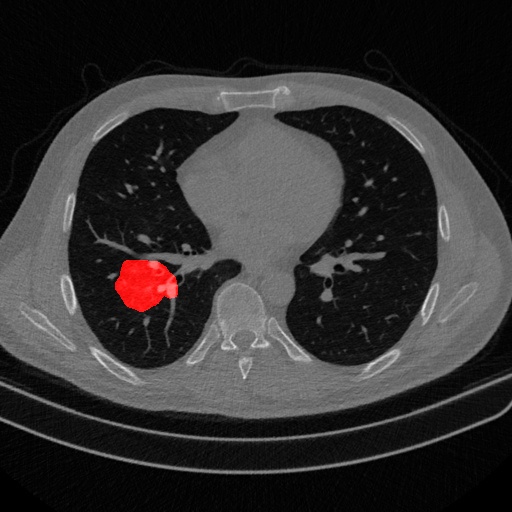}}}
\subfloat[\stargantwo]{
\resizebox*{3.5cm}{!}{\includegraphics{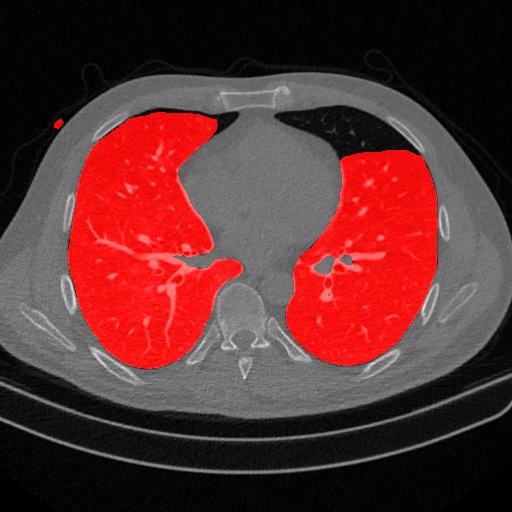}}}
\subfloat[\stargantwo + F]{
\resizebox*{3.5cm}{!}{\includegraphics{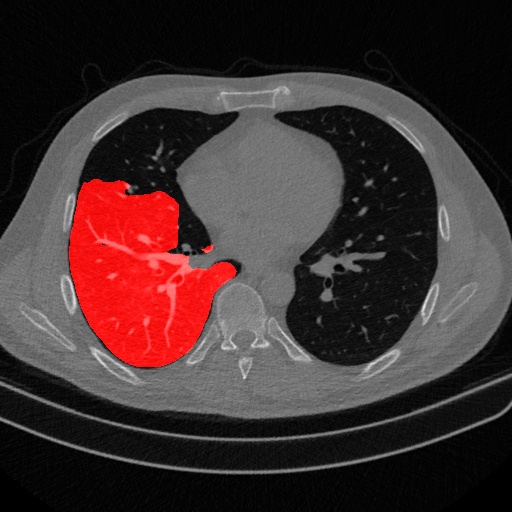}}}
\subfloat[\stylegantwoada + F]{
\resizebox*{3.5cm}{!}{\includegraphics{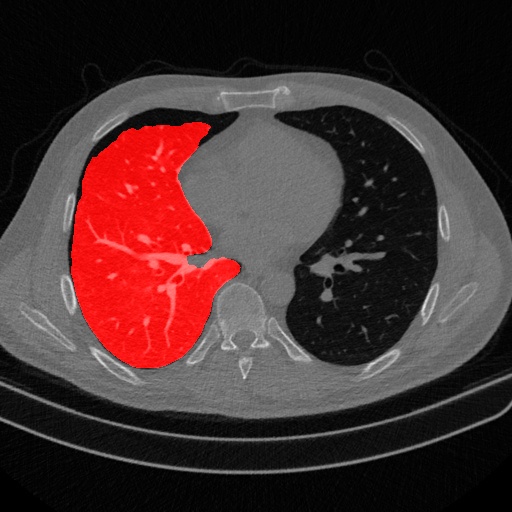}}}
\subfloat[\stylegantwoada + F]{
\resizebox*{3.5cm}{!}{\includegraphics{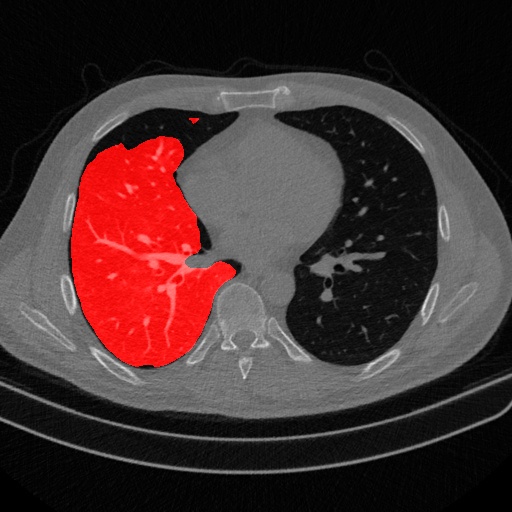}}}
}
\caption[]{Qualitative comparison between the proposed data augmentation and the baseline without data augmentation. With the proposed data augmentation, the segmentation network found the lesion region closer to the expected in the ground truth.}
\label{fig:qualitativeganimprove}
\end{figure}

Also, unlike the previously evaluated data augmentation technique, the proposed one is not applicable in an online manner due to the several steps needed to generate the new images. Thus, the proposed data augmentation method was evaluated through a technique called offline augmentation. In this technique, the images are generated before the network training, increasing the number of training images. The number of images was increased by $0.05$, $0.1$, $0.15$, $0.2$, \rev{$0.25$, and $0.30$} to preserve the same probabilities evaluated~previously. The proposed data augmentation method achieved better \fscores in three datasets: MosMed, Ricord1a, and Zenodo. Also, in Ricord1a, our method achieved the highest \fscore of all evaluated techniques \rev{in the unified training strategy} applied with probabilities \rev{$0.1$, $0.2$, and $0.3$}. The proposed data augmentation method generally achieved close results with the images generated with the \stargantwo and \stylegantwoada. Additionally, the version where the healthy lung images were flipped before receiving a lesion achieved close results to the version without~flipping.

\rev{It is important to highlight that our proposed approach produced the highest result on the dataset Ricord1a with the unified training strategy, which we previously indicated as more sensitive to color changes.
The proposed approach provides a variety of both cases: color and shape specifically adapted for this problem, although there is a greater focus on color. Our hypothesis is that inserting real lesions (a proportionally small quantity of pixels) on completely artificial backgrounds (the majority of the pixels) encourages a higher variation of both color and shape instead of just one of them, which proved more beneficial for this particular dataset than the traditional approaches that focus on only one aspect.}

\rev{Although it did not achieve the highest \fscore value, our approach improved the \fscore on 
 the MosMed and Zenodo datasets and achieved competitive results to the traditional approaches. In CC-CCII and MedSeg, our approach did not outperform the baseline without data augmentation. This occurred because the \glspl*{gan} were trained with the Ricord1b dataset, and the background regions are different from those datasets. Including background regions close to those datasets was left for future work.}

Moreover, in some cases, the new samples drastically improved the mask quality, as seen in \cref{fig:qualitativeganimprove}. In the first row, without data augmentation, the segmentation network could not even find any lesion, a failure case with a severe biological implication. In the example presented in the second row, without data augmentation, the segmentation network presents a meager recall value, predicting a significant false positive area. In the third and fourth rows, there is a low precision, with many false negative pixels, thus a low \fscore value, on an image where the lesions almost wholly cover the~lungs.

In medical problems, false negative results are generally the worst scenario, highly increasing patient risks~\citep{Woloshin2020,Wikramaratna2020,Kanji2021}. In such examples, the proposed data augmentation approach helps the network find a segmentation mask closer to the ground truth, avoiding a false positive result. Furthermore, as presented in~\cref{fig:qualitativeganimprove}, the proposed data augmentation method improved the segmentation task by finding a segmentation mask closer to the~expected.

In \cref{fig:qualitativeganimprove}, note that the four variations analyzed of our proposed technique managed to produce much better results with the same network architecture than those without data augmentation. We believe this is due to the small number of images in the original datasets. Thus, the diversity introduced by the proposed technique improved the generalization in some cases.
\section{Conclusion and Future Work}

In this work, we evaluated $24$ data augmentation techniques on COVID-19~\gls*{ct} scans on five datasets using an encoder-decoder network composed of RegNetx-002~\citep{xu2021regnet} and U-net++~\citep{Zhou2018}, comparing \rev{six} different probabilities of applying the techniques \rev{$0.05$, $0.1$, $0.15$, $0.2$, $0.25$, and $0.3$}. To the best of our knowledge, this is by far the most extensive evaluation done on this topic.
\rev{Two limitations of our work are: (i)~parameter optimization was left out; and (ii)~although six discrete probabilities were evaluated, the results indicate that there may be an optimal probability outside the range $[0.05, 0.3]$. Hence, more experiments should be conducted in future~works.}

In addition to the traditional approach, where the training and test sets are disjoint subsets from the same dataset, we proposed a different methodology where the training subsets are combined into a single larger set. We showed that the data augmentation techniques were consistently more effective (i.e., they reached better \fscore and \gls*{iou} values) with this training strategy in four of the datasets we performed experiments. \rev{The Ricord1a was the only dataset which did not achieved improvements with the unified training set. This happened due to the balancing approach that prioritized the small datasets that achieved poor results in the traditional training strategy when compared with the Ricord1a}. Also, although applying data augmentation techniques with a probability of \rev{$0.3$} in the first evaluation did not show interesting results, the same probability produced the highest number of data augmentation techniques that improved the \fscore when the datasets were combined into a single training set. Furthermore, the slight difference achieved by the data augmentation process in the individual training sets resulted in a small difference between the probabilities applied, making it unclear which is the optimal probability in this case. Meanwhile, applying data augmentation in the unified training set achieved overall higher results, clarifying that increasing the probability of the data augmentation techniques generates better results, \rev{opposing previously works from literature~\citep{muller2020automated, Mller2021}}. Lastly, by using a higher image resolution than our previous work~\citep{krinski2022} we obtained a significant gain in \fscore in four of the datasets~evaluated.

Additionally, our results show that the best operations in both training strategies are those that change shape instead of colors, such as Grid Distortion, Optical Distortion, Flip, Piecewise Affine, and Shift Scale~Rotate. According to our results, data augmentation techniques like \gls*{clahe}, Coarse and Grid Dropout, Random Crop, Image Compression, Random Gamma, and Random Snow did not generate \rev{improved} results and thus do not need to be applied to this~problem.

Finally, we proposed a novel data augmentation technique that first employs a \gls*{gan} model to produce new \gls*{ct} scans of healthy lungs and then combines existing labeled lesions with those new images to generate new samples and boost the segmentation performance. According to our results, this technique is promising, managing to improve the segmentation in certain cases. \rev{The \stargantwo + F with a probability $0.3$ achieved the highest \fscore on the Ricord1a dataset in the unified training strategy.} However, there is still room for a more consistent improvement in future works. The generated images are similar to the real images with a low \gls*{fid}, and the trained segmentation models managed to generate a reasonable ground truth annotation for those new samples. This presents an opportunity for future works to improve our proposed data augmentation focusing on the COVID-19 \gls*{ct} segmentation problem but also opens the possibility of applying our pipeline to other medical problems with lesions inside the lung~regions. The code used to perform our experiments is publicly available at \supplementary.
\section*{Acknowledgments}
We thank the Coordination for the Improvement of Higher Education Personnel (CAPES) for granting a PhD scholarship to two of the authors. We also thank the National Council for Scientific and Technological Development (CNPq) for funding the second author. We gratefully acknowledge the support of NVIDIA Corporation with the donation of the GPUs used for this research, as well as the C3SL-UFPR group for the computational cluster~infrastructure.
\section*{Disclosure statement}
No potential conflict of interest was reported by the author(s).

\section*{Notes on contributors}
Bruno Krinski is a PhD student at the Federal University of Paraná (UFPR), where he also received his master's degree in Computer Science (2019). He received his bachelor's degree in Computer Science (2016) from the UFPR.  His research interests include deep learning, image processing, and computer vision.

Daniel V. Ruiz is a Master of Science (M.S) in Informatics at the Federal University of Paraná (UFPR) (2021). B.S. Computer Science at UFPR (2017). Exchange student via the CAPES-UNIBRAL program, an exchange program associated with the program Science without Borders, at the Friedrich-Alexander-Universität Erlangen-Nürnberg (2015-2016). Working fields: robotics, computer vision, machine learning, computer graphics, digital image processing, and web~development.

Rayson Laroca received the bachelor’s degree in software engineering from the State University of Ponta Grossa (UEPG), Brazil, and the master’s degree in computer science from the Federal University of Paraná (UFPR), Brazil. He is currently a PhD student with UFPR. His research interests include computer vision, machine learning, and pattern recognition.

Eduardo Todt is an Electrical Engineering (1985) and Master in Computer Science (1990) from the Federal University of Rio Grande do Sul (UFRGS) and PhD in Advanced Automation and Robotics - Polytechnic University of Catalonia (IRI-UPC, 2005), in partnership with FZI Institute - Karlsruhe, obtaining the title of European Doctor, with Honor. Since 2008, professor at the Department of Informatics of the Federal University of Paraná (UFPR), coordinator of the Computer Science course from 2010 to 2014 and vice-coordinator of the Biomedical Informatics since january 2021. Coordinator of large scale projects related to Educational Linux, Educational Objects Platform, and dahsboard of the National Program of Books and Didatical Material (PNLD). Member of research groups Center of Scientific Computation and Free Software (C3SL), Vision, Robotics and Images (VRI), and Sustainable Traffic and Transport (TTS) at UFPR. Experience in Electrical Engineering and Computer Engineering, developing products and systems for industrial automation, as well as R\&D manager. Main interests are on mobile robotics, computational vision and computer science in education. Member of the Steering Committee of the Special Commission on Robotics (CER) of the Brazilian Computer Society (SBC) since 2018.

\bibliographystyle{tfcse}
\bibliography{main}

\end{document}